# Optical to infrared mapping of vapor-to-liquid phase change dynamics using generative machine learning

Siavash Khodakarami[1,2,*], Pouya Kabirzadeh[1], Chi Wang[1,3],
[*,]Tarandeep Singh Thukral[1], Nenad Miljkovic[1,4,5,6,7,8]

[1] *Department of Mechanical Science and Engineering, University of Illinois at Urbana – Champaign, Urbana, Illinois 61801 USA*

[2] *Current Address: Division of Applied Mathematics, Brown University, Providence, RI 02912 USA*

[3] *Current Address: Department of Mechanical Engineering, University of New Mexico, Albuquerque, NM 87131, USA*

[4] *Department of Electrical and Computer Engineering, University of Illinois at Urbana – Champaign, Urbana, Illinois 61810 USA*

[5] *Material Research Laboratory, University of Illinois at Urbana – Champaign, Urbana, Illinois 61801 USA*

[6] *Air Conditioning and Refrigeration Center, University of Illinois at Urbana – Champaign, Urbana, Illinois 61801 USA*

[7] *Institute for Sustainability, Energy and Environment (iSEE), University of Illinois, Urbana, Illinois, United States*

[8] *International Institute for Carbon Neutral Energy Research (WPI-I2CNER), Kyushu University, 744 Motooka, Nishi-ku, Fukuoka 819-0395, Japan*

**\*** Corresponding Authors Emails: Siavash_khodakarami@brown.edu, nmiljkov@illinois.edu

**Abstract**

Infrared thermography is a powerful tool for studying liquid-to-vapor phase change processes. However, its application has been limited in the study of vapor-to-liquid phase transitions due to the presence of complex liquid dynamics, multiple phases within the same field of view, and experimental difficulty. Here, we develop a calibration framework which is capable to studying one of the most complex two-phase heat transfer processes: dropwise condensation. The framework accounts for non-uniformities arising from dynamic two-phase interactions such as droplet nucleation, growth, coalescence, and departure, as well as substrate effects particularly observed on micro- and nanoengineered surfaces. This approach enables high-resolution temperature measurements with both spatial (12 μm) and temporal (5 ms) precision, leading to the discovery of local temperature phenomena unobservable using conventional approaches. These observed temperature variations are linked to droplet statistics, showing how different regions contribute to local condensation heat transfer. We extend the developed method to quantify local thermal parameters by fusing it with a generative machine learning model to map visual images into temperature fields. The model is informed of the physical parameter by incorporating vapor pressure embedding as the conditional parameter. This work represents a significant step toward simplifying local temperature measurements for vapor-to-liquid phase change phenomena by developing a methodology as well as a machine learning approach to map local thermal phenomena using only optical images as the input.

**Introduction**

Phase change heat transfer is essential for electrification and decarbonization of society. This includes electrified vehicles, next-generation electronics, and data center thermal management for power hungry artificial intelligence tasks. Fundamental understanding of vapor-to-liquid phase change phenomena requires high-resolution temperature measurements which can improve both understanding and development of physics-informed models to enable enhanced heat transfer surface design to meet the growing demands of industry. However, local heat transfer measurement during vapor-to-liquid phase change processes is challenging. Most studies rely on attaching several temperature sensors such as thermocouples, resistance temperature detectors, or thermistors locally at different locations on the heat transfer surface, obtaining the spatially-averaged local temperature and heat flux with relatively low spatial resolution.[1,2] This method of sensor use is mainly applicable to internal two-phase flow scenarios such as flow condensation and evaporation where the temperature sensors are attached to the outer wall of the tube or channel of interest. To determine the internal heat transfer dynamics, the internal wall temperature is estimated using the external temperature measurement in combination with the wall thermal resistance.[3] The main challenges with established thermometry methods are the: necessity to use numerous thermocouples carefully attached on the heat transfer surface to obtain a reasonable number of local measurements, and low-spatial resolution and discrete measurement due to the limited number of temperature sensors which can be attached to the surface of interest.

For vapor-to-liquid phase transitions on external surfaces[4,5], use of discrete temperature sensors for local measurement presents even greater challenges. Thermal sensors act as pinning points for liquid motion, interfering with the two-phase phenomena of interest. The presence of sensors on the surface ensures a difference in fluid dynamics compared to a sensor-free surface.

Whether internal or external, one of the biggest challenges to spatially discrete thermometry is the attachment method of the sensor to the surface. To accurately obtain a true measurement of the surface location of interest, an experimentalist must take extreme care in understanding the thermal impedance that arises due to the attachment method used.[6] For example, soldering, adhesive bonding, or thermal pastes all create additional thermal resistances that impede the ability to obtain a true measurement, and which are exacerbated by the typically low temperature differences between the surface and fluid characteristic of two-phase heat transfer processes[7]. This in turn drastically increases measurement uncertainty, and reduces spatial fidelity in the heat transfer measurement.[8]

To grapple with the difficulties associated with discrete thermometry techniques, non-intrusive measurement methods such as infrared (IR) thermography are required and have been investigated in the past.[9-11] Previous studies have taken advantage of IR thermography for local measurement in liquid-to-vapor phase change, enabling local heat flux measurements during boiling processes.[10, 12-14] IR thermography has been also used to study both the onset and enhancement of critical heat flux (CHF) in boiling.[10, 15-18]

While IR thermography has been successfully utilized for local heat transfer analysis during boiling, only a handful of studies have applied these techniques to steam condensation in the absence of non-condensable gases (NCGs). For pure steam condensation tests, a test section should be installed inside a vacuum compatible chamber to remove NCGs. Hence, the IR camera can only obtain the IR signal through a viewport. Furthermore, common condensing surfaces are metallic (e.g., copper and aluminum), with low emissivity and high reflectivity. Unlike in boiling studies, backside IR imaging is not feasible during condensation on these metallic surfaces. As a result, it is essential to account for both the reflections from inside and outside of the condensation chamber, and the viewport interference.

Previous studies using IR thermography for condensation have faced limitations such as low spatial or temporal measurement resolutions, focusing on flat surfaces instead of tubes and not examining pure steam condensation devoid of NCGs.[11, 19, 20] Several studies have focused on studying single droplet phase change using IR thermography, which requires an extension to actual condensation processes involving multiple droplet interactions.[21-24]

Infrared thermography enables non-intrusive high resolution temperature measurements for phase change studies, enabling fundamental advancements in understanding of complex and stochastic heat and mass transfer. It also paves the way for the development of precise models, which can potentially lead to enhanced design of the heat transfer surface. However, high cost and complex calibration methods have limited its use in the study of phase change phenomena. In contrast, optical imaging is more common due to its simplicity, lower cost, and simplified calibration requirements. With the recent advancements in artificial intelligence (AI), vision-based models can extract physical features or predict averaged heat transfer from optical data.[25-27] However, they lack the ability to provide localized temperature measurements. Generative AI presents an opportunity to fill this gap by enabling the mapping of optical images into temperature fields. In recent years, generative adversarial networks[28] (GANs) have been used to predict temperature fields through IR-like image generation in other scientific applications such as sea surface temperature[29], computer chip thermal maps[30], forest fire monitoring[31], and flow temperature during mixed convection in channels[32]. Previous studies have mainly focused on generating IR-like images from visible images (e.g., RGB) with no emphasis on the underlying physics. A caveat is that these models do not prioritize accurate temperature prediction, although effective in image generation.

In this work, we resolve the challenges by developing a machine learning method to accurately map images into high spatial and temporal resolution thermography profiles for a complex two-phase heat transfer phenomenon. We first used high-speed IR imaging to study the transient temperature distribution during pure steam DWC on tubular surfaces with high temporal and spatial resolution. A robust pixelwise calibration framework is developed that allows for temperature estimation from IR signals taken from highly reflective metallic tubes having low emissivity during DWC. Using the developed method, a two-step calibration is used to separate the solid substrate and condensate droplets, enabling for the use of IR thermography to investigate temperature distributions and track phase change processes with high accuracy and resolution. Despite its potential, the developed method is hindered by its requirement for time-intensive calibration, data collection, post-processing, and use of a high cost IR camera. To ease the burden of these requirements, we developed a generative model based on GANs to map gray-scale images into temperature fields during DWC. We show that this model can effectively predict temperature variation of condensate droplets and the solid substrate at different vapor pressures. The model is conditioned by the vapor pressure, enabling it to accurately generate temperature fields that correspond to the appropriate range based on the given vapor pressure. The model is conditioned on both input images and the vapor pressure parameter, making it more comprehensive than known conditional GANs (cGAN[33]). Our work develops a previously unexplored method to simplify the acquisition of accurate temperature maps during two-phase condensation phenomena by developing a machine learning approach which can predict thermal phenomena using only gray-scale images as an input. Our work is a powerful approach which can be extended to a variety of other two-phase phenomena for developing mechanistic understanding.

**Results**

**Condensation infrared thermography and calibration**

Pure steam condensation data are collected from experiments conducted with the vacuum compatible facility depicted in Fig. 1. Extensive details regarding this experimental facility are available in prior works (see Methods).[25, 34] IR imaging data as well as temperature and pressure sensor signals were collected during these experiments. The total radiation measured by the IR camera includes radiation emitted from the condenser tube, radiation emitted from the sapphire viewport, atmospheric radiation, reflection of the surrounding environment inside the chamber (e.g., chamber walls), and the background reflection (Fig. 1a). Since the chamber is initially evacuated of all air and NCGs and pure steam condensation tests are conducted at low pressures (< 10 kPa), atmospheric radiation from inside the chamber could be neglected. The total spectral radiation flux measured by the camera is a sum of the contributions from the viewport, test tube, chamber wall, and background environment temperatures, as well as the optical properties in the working wavelength range (e.g., 3 μm to 5 μm) (see Supplementary Note 1).

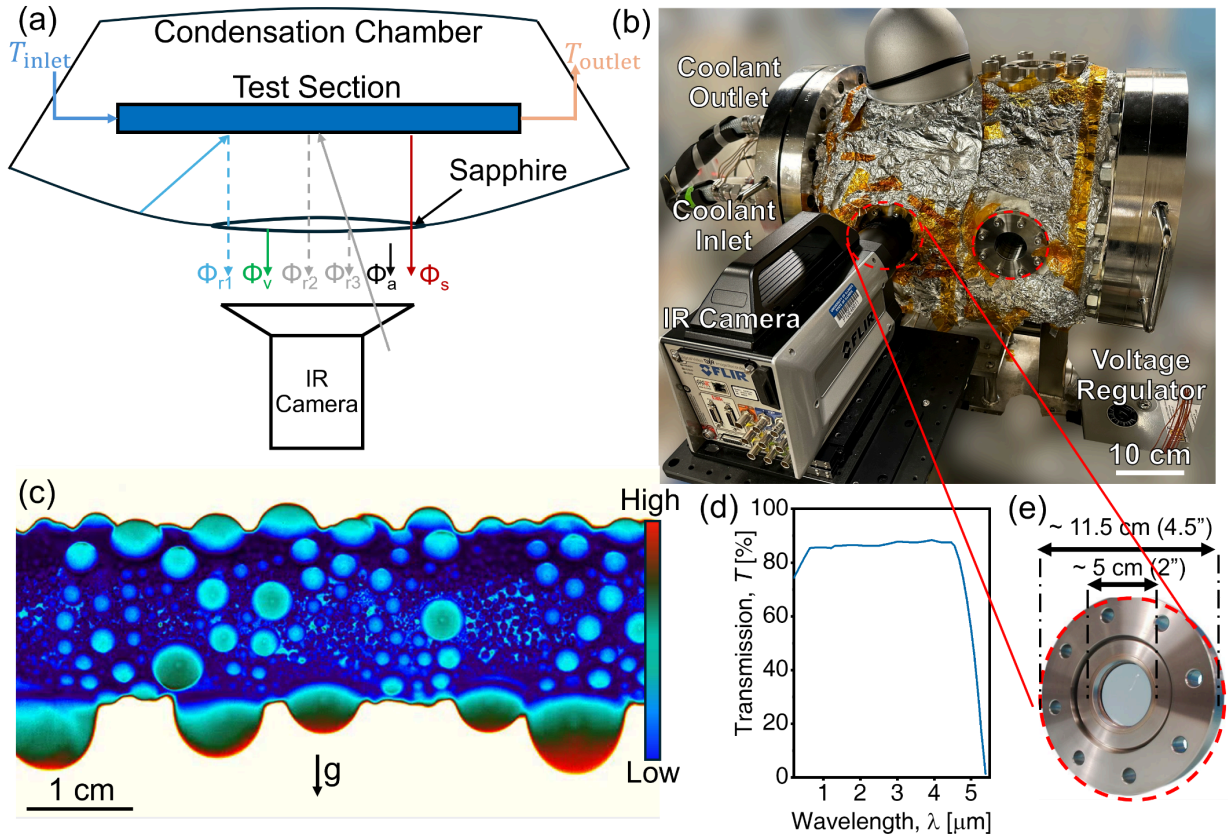

**Figure 1. Experimental method and exemplary data.** (a) Schematic showing the radiation measured by the IR camera. Schematic not to scale. (b) Photograph of the IR camera positioned for IR imaging through the chamber viewport. The background is artificially blurred to provide focus on the experimental setup only. (c) Example IR image obtained through the viewport (not calibrated). (d) Measured transmission ($T$) of the sapphire glass viewport as a function of wavelength ($\lambda$). (e) Photograph of the sapphire glass viewport with specified dimensions.

All optical properties need to be defined to correctly convert total measured radiation into condenser surface temperature. Although challenging, this method is feasible for uniform surfaces with a constant emissivity. However, many micro/nanostructured surfaces and coatings lack uniformity, and emissivity could fluctuate across the surface in different directions. For example, a copper oxide (CuO) layer fabricated on a copper tube shows pixel-to-pixel emissivity variation. Therefore, assuming a constant emissivity value leads to erroneous temperature estimation. Additionally, both solid surface and condensate droplets with varying emissivity simultaneously coexist on the surface during DWC. This necessitates a tailored approach to accurately account for the differing radiation properties of the droplet and solid regions. Here, we developed a two-step calibration framework that can alleviate the challenges associated with testing reflective surfaces with low emissivity (e.g., metallic surfaces), non-uniform coatings, and coexistence of metallic solid substrate and condensate droplets.

Prior to each condensation test, the condenser surface temperature was varied in the range of 7°C to 40°C by varying the coolant temperature based on the chiller set point, covering the surface temperature range during the condensation test. The coolant flow rate was set to maximum during the calibration step (~25 liters per minute (LPM)) to ensure temperature uniformity. Surface temperature was calculated based on an energy balance at the solid-air interface. At every surface temperature, raw IR signals or counts were recorded with the camera for a few seconds at a frame rate of 181 fps which is the maximum value at the camera full resolution of 1280 × 1024 pixels. Temporally averaged temperature was used as the representative pixel-wise calibration matrix for each temperature. The final substrate calibration matrix had a size of ($N_T$, $H$, $W$), where $N_T$ is the number of temperature steps that data has been recorded, and $H$ and $W$ were the spatial resolutions. The typical number of temperature steps was 8 to 10. During condensation, vapor pressure varied from 2 kPa to 8 kPa. At every vapor pressure, IR counts were measured for 11 seconds at full resolution and recorded as a matrix of size ($N_t$, $H$, $W$), where $N_t$ denotes the number of time-steps. Temperature distribution could be estimated by interpolation from the substrate calibration matrix at each pixel and each time-step. However, condensate droplets with different emissive properties make the substrate calibration matrix inapplicable to droplet regions. Additionally, droplets form, grow, and move with different rates depending on the vapor pressure and surface properties (e.g., wettability), necessitating a separate calibration scheme for condensate droplets. We assume that the tip of the hanging droplets (Fig. 2) corresponds to the vapor saturation temperature due to the large thermal resistance across large droplets, which minimizes the heat transfer at the tip, ensuring that its temperature closely matches the saturation temperature.[35]

To accurately convert raw IR signals into temperatures, we defined key parameters such as emissivity, reflected temperature, and external optics settings in the camera software to ensure that the temperature at the tip of the large hanging droplets matches the saturation temperature at every vapor pressure. Condensate droplets were assumed to exhibit uniform reflectivity due to their high emissivity (> 0.95), and the small field of view (~ 1 cm). We validated this assumption by setting the parameters based on one droplet tip temperature and observing consistent temperatures across other large droplets residing at different regions.

After generating substrate and droplet calibration matrices, each time-step from the condensation IR video was processed by a droplet segmentation model (see Methods and Supplementary Note 2) to locate the exact location of the droplets. At each time-step, droplet-covered regions were detected, and the droplet temperatures were derived from the droplet calibration matrix, while substrate temperatures were interpolated from the substrate matrix. This approach ensured accurate temperature estimation for both substrate and droplet and eliminated the uniform emissivity assumption. Summary of the calibration framework is shown in Fig. 2. The IR imaging system used in this study achieves fine spatial and temporal resolution of ~12 μm, and ~5 ms, respectively.

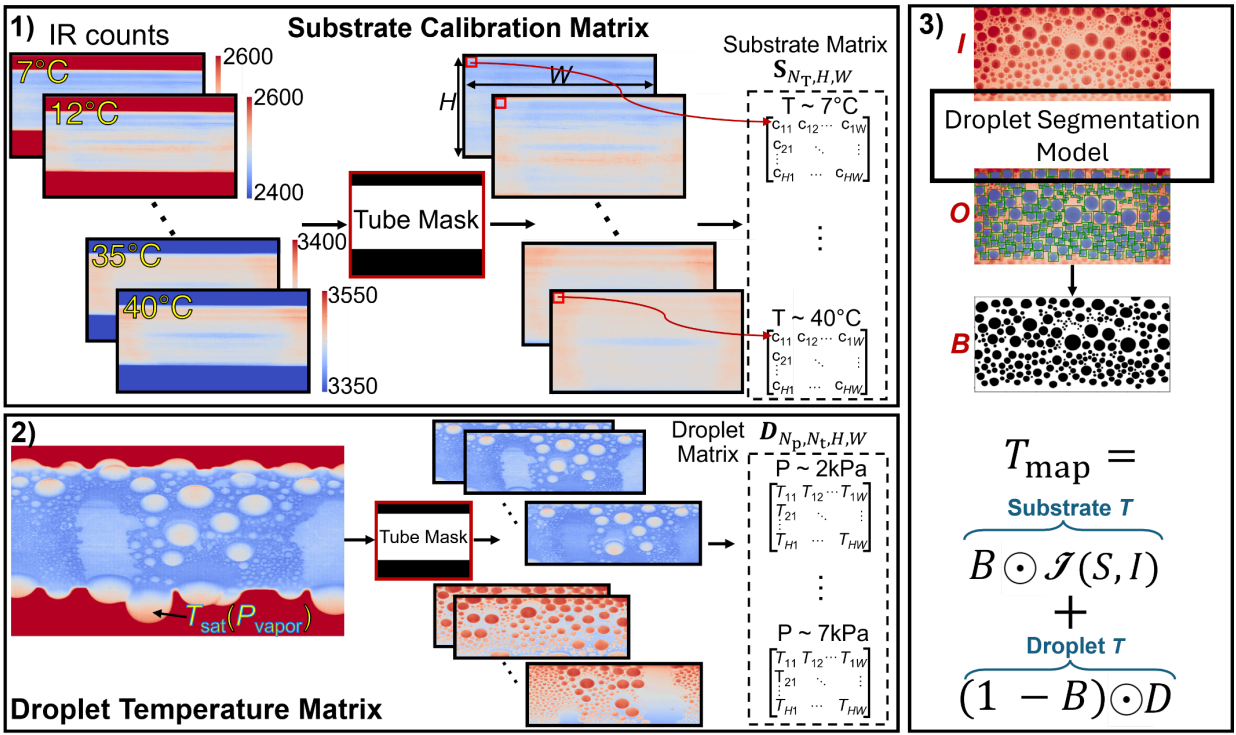

**Figure 2. Summary of the pixelwise calibration methodology developed for IR imaging of dropwise condensation experiments**. The first module (labled 1) is conducted prior to condensation experiments to obtain the substrate calibration matrix. The second module (labeled 2) is used for obtaining the droplet temperature matrix which is only valid in areas where droplets exist (and not in areas where the substrate is exposed). The third module (labeled 3) is

used to detect and segment droplets, create a binary map of droplets and substrate, and estimate the final temperature map at each time step using both the substrate matrix (*S*) (from module 1), and droplet temperature matrix (*D*) (from module 2). Here, $N_T$, $N_t$, and $N_p$ are the number of temperature steps in the substrate calibration matrix, number of time-steps, and number of vapor pressures, respectively. I is the interpolation operator.

**Temperature distribution during dropwise condensation**

To investigate the application of IR imaging and the developed calibration framework, we fabricated several hydrophobic tubes and a superhydrophobic tube to study the droplet dynamics and temperature distributions during DWC and jumping droplet condensation at various vapor pressures. The surfaces included copper oxide (CuO) with a conformal 2 μm thick Parylene C coating, CuO with 1 μm thick polydimethylsiloxane (PDMS) layer, CuO with a 5 μm thick Gentoo layer, Cu with a 1 μm thick PDMS layer, and a nanostructured additively manufactured (AM) Al-alloy coated with Heptadecafluoro-1,1,2,2-tetrahydrodecyl trimethoxysilane (HTMS) to achieve superhydrophobicity. More details on AM Al-alloy nanostructuring can be found in previous work[36]. All fabrication steps are described in Supplementary Note 3.

The micro-structured CuO surfaces were fabricated through oxidation in a strong basic solution at high temperatures, resulting in needle-like microstructures and visually black color surface[37] (see Supplementary Fig. 1). The black color reduces the reflectivity of Cu surface from more than 0.9 to the range of 0.35 to 0.7 in the wavelength range of 3 μm to 5 μm (see Supplementary Fig. 1).[38] CuO is intrinsically superhydrophilic, resulting in droplet contact angle smaller than 5°. Hydrophobic coatings were deposited on the CuO structures to promote DWC. Droplet contact angle measurements ensured advancing contact angle larger than 90° for all the samples (see Supplementary Table 3 and Supplementary Note 4).

Examples of dynamic 2D temperature maps during DWC are shown in Fig. 3. It is evident that condensate droplets are at higher temperatures compared to the substrate areas, with larger droplets exhibiting higher temperatures compared to smaller droplets. This is due to the lower thermal resistance across the smaller droplets, allowing for more effective cooling of the droplet from the bottom.[35, 39] On the other hand, very large droplets remain at the saturation temperature corresponding to the vapor pressure. Also, the rate of droplet growth increases with vapor pressure. As shown in Fig. 3, the overall temporal temperature fluctuation is higher at higher vapor pressure. This is due to the accelerated droplet cycle including nucleation, growth, and shedding or sweeping at higher vapor pressures.[25, 40]

Droplet nucleation happens at the saturation temperature, but very small droplets are rapidly cooled down by the cold surface. As droplets grow, their thermal resistance across the droplet increases, resulting in a larger temperature difference across them. Once a droplet reaches a critical diameter where the gravitational forces applied to the droplet exceed the droplet-substrate adhesion forces, the droplet starts moving down in the gravity direction. This results in a sudden temperature drop on the surface as the hot droplet is removed and the cold substrate is exposed to the camera. Furthermore, large moving droplets may sweep away smaller droplets, exposing more cold areas to the camera. An example of this can be seen between $t = 0$ and $t = 50$ ms at vapor pressure of 4.2 kPa in Fig. 3.

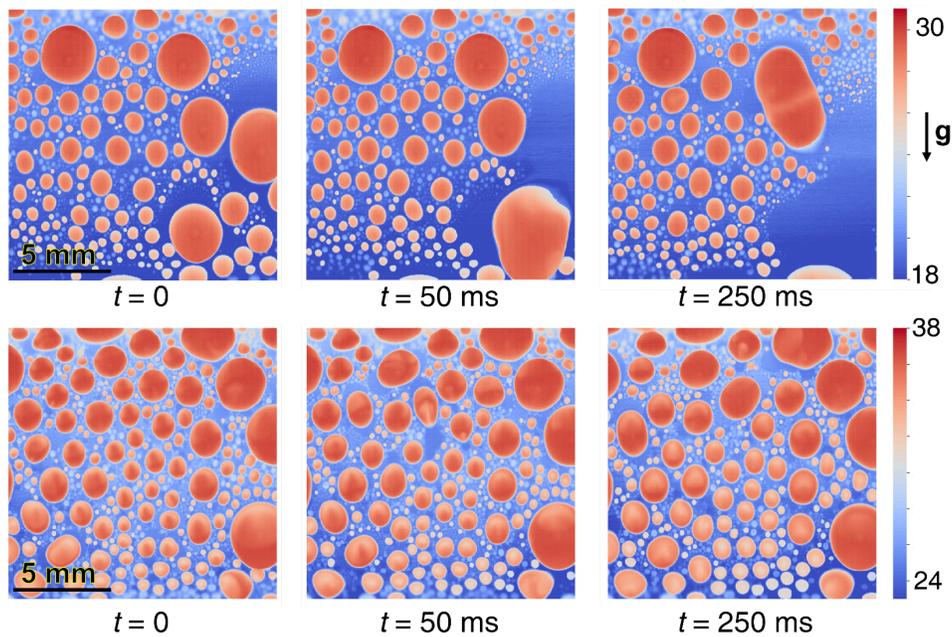

**Figure 3. Time-lapse of temperature distribution on a hydrophobic surface during condensation captured through IR thermography**. (a) Results showing temperatures during condensation at $P_{vapor} = 4.2$ kPa. (b) Results showing temperatures during condensation at $P_{vapor} = 6.2$ kPa. The hydrophobic surface in both tests was CuO with a PDMS coating. Temperatures are in units of $^0$C.

We observed that droplets in the top region of the tube remain pinned to the surface for longer durations compared to those in the middle or bottom areas (gravity being from top to bottom). This is due to the smaller gravitational force component in the surface tangential direction imposed on droplets sitting on the upper region of the tube. Note that tube edges are not included in this analysis, and only the area between the edges is considered (see Fig. 4). Due to the extended retention of droplets in the top region, the time-averaged temperature is higher at the top of the tube and decreases in the direction of gravity. To quantify this trend, we plotted the vertical (y-direction) temperature variation from top to bottom of the tube at different axial locations (x-direction) in Fig. 4. These results are based on time-averaged temperature measurements at each vapor pressure. Five randomly selected lines were analyzed to make sure that the obtained results were not due to a defect in a specific region of the tube or just random variations. The temperature variation across all five lines followed a consistent trend. As shown in Fig. 4b, the time-averaged vertical temperature variation was negligible at very low pressures. However, a temperature difference of approximately 3°C to 4°C was observed at higher vapor pressures. It should be noted that the temperature gradient was steeper near the top and bottom areas of the tube and was minimum in the middle part of the tube.

To validate this observation, we conducted the same analysis on condensation IR thermography data of the CuO with PDMS sample (Supplementary Fig. 2). Similarly, there was no significant temperature decrease from top to bottom of the tube at vapor pressure of 2.3 kPa. However, there was a clear decreasing trend in the time-averaged temperature from top to bottom of the tube at higher vapor pressures. The temperature dropped more sharply in the top area of the tube, specifically in the top quarter of the tube, and the smallest gradient generally happened in the middle area of the tube.

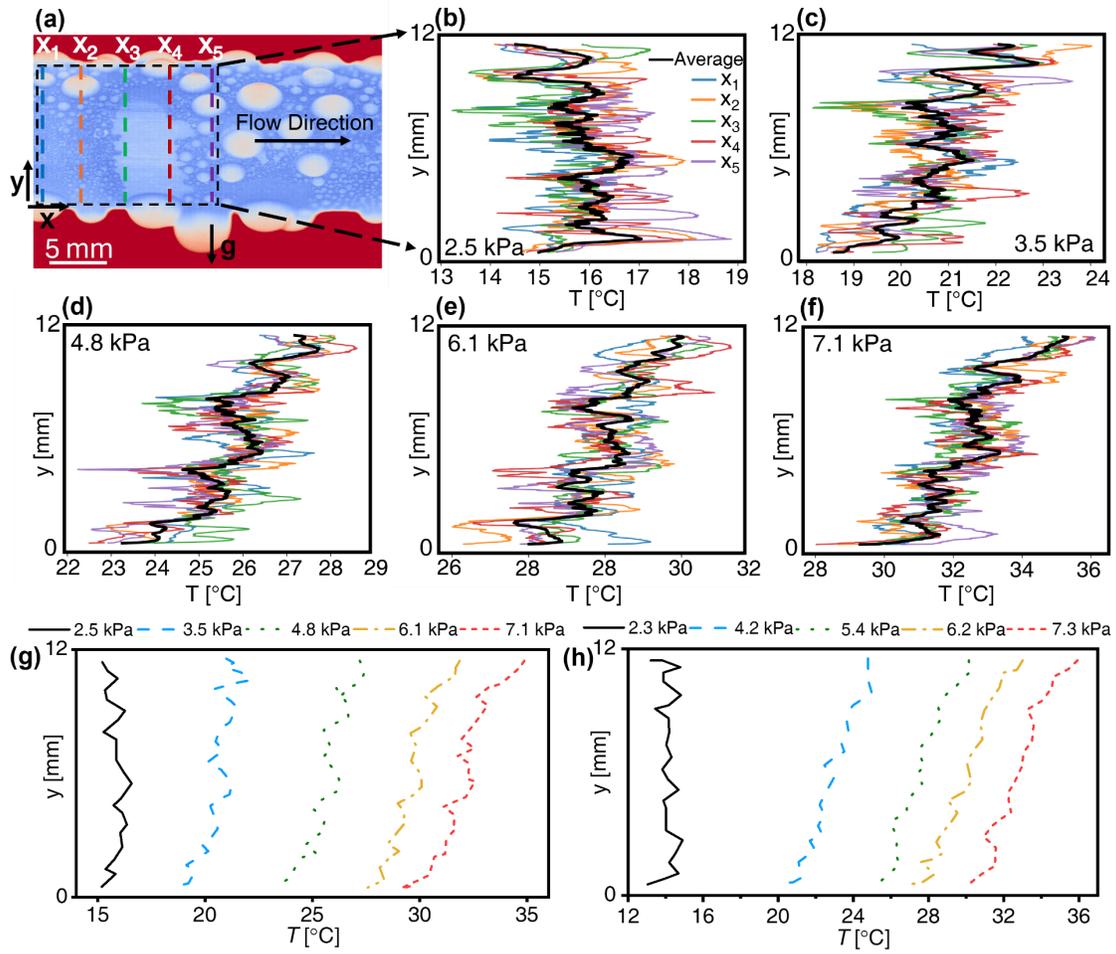

**Figure 4. Results showing the time-averaged temperature variation in the vertical direction (direction of gravity) during dropwise condensation.** (a) Image showing the analyzed area with vertical lines ($x_1$ to $x_5$) used to visualize temperature variation and the x and y coordinates. Time-averaged vertical temperature variation at vapor pressure of (b) $P_{vapor}$ = 2.5 kPa, (c) $P_{vapor}$ = 3.5 kPa, (d) $P_{vapor}$ = 4.8 kPa, (e) $P_{vapor}$ = 6.1 kPa, and (f) $P_{vapor}$ = 7.1 kPa. The legend in (b) is applicable to (c-f). The results in (b-f) are based on condensation on the Cu tube structured with CuO and coated with Parylene C. The black thicker lines in (b-f) demonstrate average temperatures of the five colored lines at $x_1$ to $x_5$. Time and axially (x-direction) averaged temperature variation on a (g) Cu tube structured with CuO and coated with Parylene C, (h) and Cu tube structured with CuO and coated with PDMS. For all experiments: tube diameter is 12.7 mm (0.5"). The analyzed area has a surface area of approximately 12 mm (height) × 14.7 mm (width).

To analyze temperature variation along the y-direction with vapor pressure, we computed the time and x-direction averaged temperature profiles, yielding a representative temperature line at each pressure (Figs. 4g and 4h). These results are presented for CuO with Parylene C and CuO with PDMS coatings. The results indicate that temperature variations are similar during DWC on hydrophobic tubes with comparable wettability characteristics (e.g., surface roughness and droplet contact angle). Temperature variation is minimal at low vapor pressures. However, as vapor pressure increases, the average temperature varies by 2°C to 6°C from top to bottom of the tube depending on the vapor pressure. The maximum temperature difference, typically between the near top and near bottom regions, increases with vapor pressure. For CuO with Parylene C tube, the maximum difference between the highest and lowest points is approximately 0.06°C, 2.5°C, and 5.7°C at vapor pressures of 2.5 kPa, 3.5 kPa, and 7.1 kPa, respectively, C. These values are 0.35°C, 4.8°C, and 5.9°C at vapor pressures of 2.3 kPa, 4.2 kPa, and 7.3 kPa, respectively, on a CuO with PDMS surface.

The temperature gradient also increases with vapor pressure. To quantify this trend, we performed linear regression using the average temperature profiles shown in Fig. 4(g-h). At a vapor pressure of ~ 2.5 kPa, the slope is almost zero on both tubes. On the CuO with Parylene C tube, the temperature increases at a 0.18 °C/mm rate from bottom to top of the tube at a vapor pressure of 3.5 kPa, increasing to 0.27 °C/mm, 0.31 °C/mm, and 0.37 °C/mm at vapor pressures of 4.7 kPa, 6.1 kPa, and 7.1 kPa, respectively. For the CuO with PDMS tube, slope is 0.37 °C/mm, 0.41 °C/mm, 0.45 °C/mm, and 0.43 °C/mm at vapor pressures of 4.2 kPa, 5.4 kPa, 6.2 kPa, and 7.3 kPa, respectively.

The temperature variation across the vertical direction on the tube is primarily due to the differences in droplet residence time. Droplets generally remain on the upper region of the tube for longer periods, allowing them to grow larger when compared to those in the middle or the bottom of the tube (excluding the edges).

To test this hypothesis, we extracted three time-dependent parameters: surface coverage, droplet count, and average droplet size across the top, middle, and bottom thirds of the tube (Fig. 5). During DWC, the top region has a larger average droplet size more frequently compared to the mid and bottom regions. Similarly, the droplet coverage is higher on the top region, followed by the mid and bottom regions. In general, droplets stay longer and grow larger on top region before shedding. However, the time-averaged number of droplets is lower on the top region, with the mid region having the most time with a larger droplet count. Despite fewer droplet count, the time-averaged droplet coverage area is higher on the top region due to extended residence times.

While these statistics may not be interpretable individually, their combination aligns with temperature trends. Larger droplet sizes, higher coverage, and slower shedding on the top region drive elevated mean temperatures. Similarly, the average droplet size and droplet coverage in the middle region is generally slightly larger than the lower region which causes an average temperature difference between these two regions. It is known for DWC that faster droplet shedding results in higher heat transfer rates. Droplet shedding clears the surface for new nucleation which is favorable for the phase change process, while areas filled with large droplets are ineffective to condensation heat transfer, akin to filmwise condensation (FWC).[41] On average, the renewal period of condensing area is longer on the top region.

In addition to DWC on hydrophobic surfaces, we also collected IR thermography data during condensation on an AM Al-alloy superhydrophobic tube. Condensation heat transfer coefficients are generally higher on superhydrophobic surfaces due to droplet removal by coalescence-induced droplet jumping.[36, 42] There are fewer pinned droplets on the surface when compared to a hydrophobic surface. In fact, most droplets are removed at very low sizes due to coalescence-induced jumping and removal.[43] As a result, the surface temperature is more uniform during condensation on the superhydrophobic surface and the droplets residence time is also much lower when compared to the hydrophobic surfaces (Supplementary Fig. 3). Lower droplet number density and smaller temperature fluctuations are noticeable by comparing with the hydrophobic tubes.

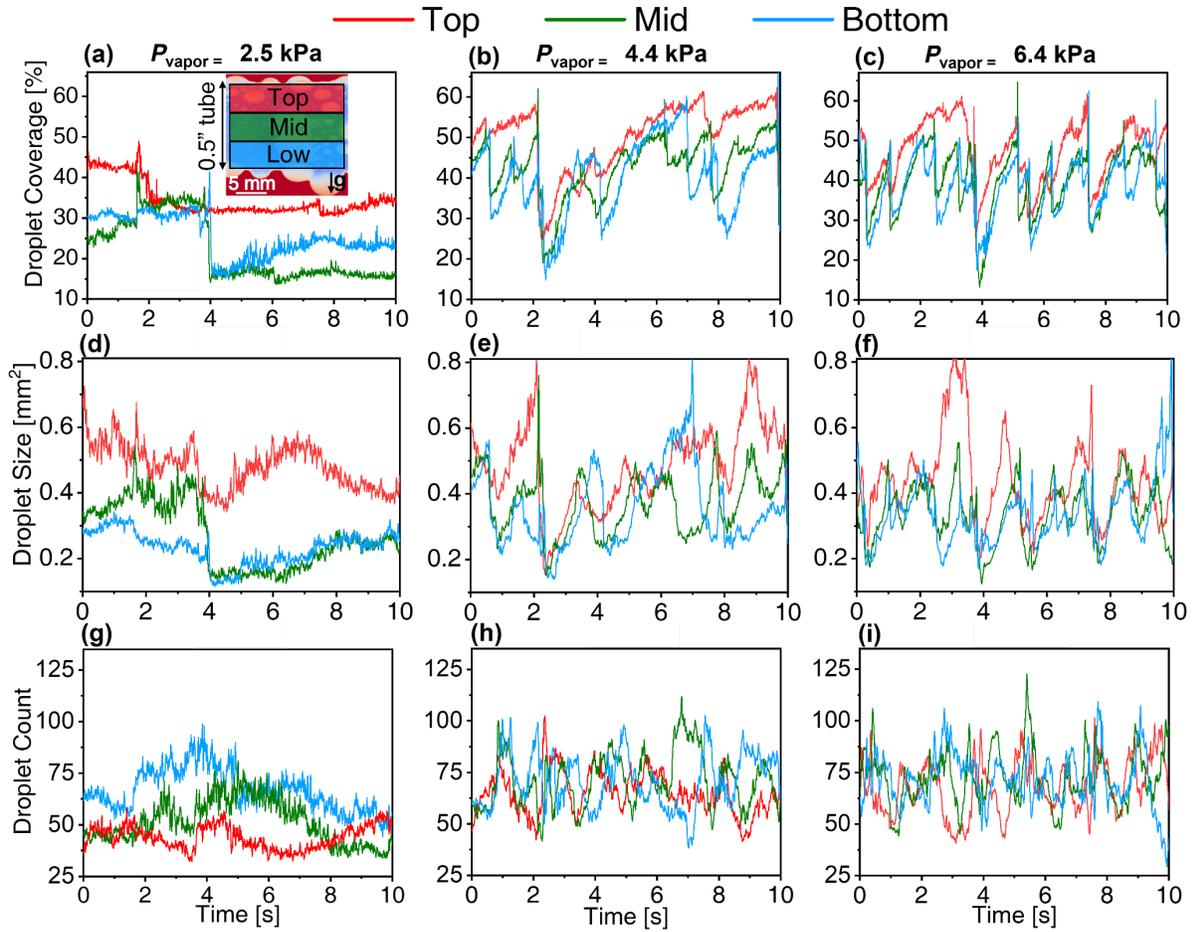

**Figure 5. Droplet statistical analysis on the CuO with Gentoo coating tube sample**.
(a- c) Transient droplet coverage (% of surface covered with droplets) at vapor pressures of $P_{vapor}$ = 2.5 kPa, 4.4 kPa, and 6.4 kPa, respectively. (e-h) Droplet size on the surface at vapor pressures of $P_{vapor}$ = 2.5 kPa, 4.4 kPa, and 6.4 kPa, respectively. (i-l) Average droplet count on the surface at vapor pressures of $P_{vapor}$ = 2.5 kPa, 4.4 kPa, and 6.4 kPa, respectively. All the analyses are performed separately on the first third, second third, and final third of the tube labeled as Bottom, Mid, and Top (top to bottom in the gravity direction) during DWC. Inset in (a) demonstrates the location of analysis on the tube.

**Condensation optical image to temperature mapping**

Access to high spatial and temporal temperature distributions during condensation is highly beneficial for fundamental phase change studies and robust modeling. However, IR imaging of pure steam condensation presents significant challenges, requiring meticulous calibration methods which are typically time-consuming and require testing in a controlled environment. Any deviation in calibration can result in significant errors. Furthermore, IR cameras capable of high-speed imaging are often expensive and not always accessible. Therefore, alternative approaches are highly desirable. To achieve this, we integrated IR imaging data with regular 2D gray-scale images to discover the mapping between image features and temperature distributions during the condensation process. Using this method, we successfully estimated temperature distributions from droplet distribution patterns and vapor pressure information. We employed conditional generative modeling informed by physics to learn the mapping between grayscale images and temperature distributions. The fusion of generative models with physical constraint enhanced the model ability to reliably generalize the temperature prediction across different vapor pressures.

We developed a conditional generative model based on GANs,[44] which consist of two neural networks, the generator $G$ and the discriminator $D$ that are trained adversarially. The $G$ learns the underlying data distribution and generates synthetic samples that are indistinguishable from real data, while the $D$ distinguishes the real data from the synthetic samples. Unlike the original GAN, which generates samples from random noise, our model, inspired by pix2pix model[45] ,uses paired IR and grayscale images with vapor pressure as conditioning variable to predict temperature distributions during DWC. Given the importance of temperature prediction accuracy, we incorporated an $L_1$ loss term to penalize pointwise temperature deviations. This loss encourages the generator to create temperatures that closely match the ground truth in addition to matching the underlying distribution through adversarial loss.

2D images of condensation were generated by discarding the temperature information and keeping the pixel intensity maps. The visual structure of the IR images remains consistent at different vapor pressures, with larger droplets being hotter than smaller ones and both being hotter than the substrate. However, the temperature range varies significantly with the vapor pressure. This problem can be reformulated through training a single $G$ that learns the mapping to multiple domains determined by the vapor pressure. However, here, the vapor pressure is not a discrete categorical label, but a continuous variable in the range specified by the experimental conditions. Given the significance of vapor pressure on the temperature range, we imposed an auxiliary regressor at the output of the $D$ to predict the vapor pressure, resulting in an additional vapor pressure loss term between the for both real and synthetic samples. A key advantage of our model is the incorporation of vapor pressure as a conditioning variable, enabling the model to generate temperature maps depending on the vapor pressure. This physics-informed approach of blending adversarial loss ($L_{cGAN}(G, D)$), temperature loss ($L_{L1}(G)$), and vapor pressure loss ($L_p(G)$, ($L_p(D)$)), enables accurate temperature mapping across vapor pressures.

A schematic of our conditional generative model is shown in Fig. 6. To effectively fuse the images with the scalar pressure value, we encoded the pressure as a fixed embedding format with two non-zero elements. The pressure embedding was constructed as follows:

A) For each vapor pressure $p_{vapor}$ (ranging between 2kPa to 8 kPa), the largest integer $n$ is identified such that $n \leq p_{vapor}$.

B) The fractional part $f = p_{vapor} - n$ is computed.

C) The pressure embedding vector $P$ is constructed such that: $P_{n-1} = 1$, $P_n = f/(n + 1)$, and all other elements are zero.

For example, a vapor pressure of $p_{vapor} = 2.5$ kPa is represented as [1,0.5/3,0,0,0,0,0]. We found that an embedding-based representation outperforms a scalar format, consistent with prior works in multi-domain image-to-image translation.[46] Here, vapor pressure serves as the physical parameter, guides the generator to map temperature distributions effectively across different domains.

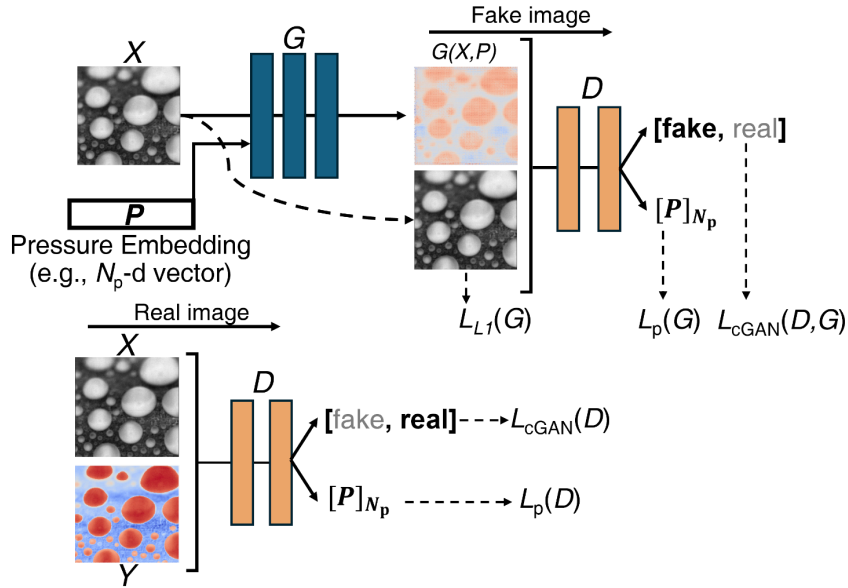

**Figure 6. Schematic of the developed pressure-informed cGAN model**. Grayscale image $X$ and vapor pressure embedding are used by the generator $G$ to create fake temperature fields. Paired $X$ and fake temperature fields $G(X,P)$ are used by the discriminator $D$ to predict the temperature field label (real or fake) and predict the vapor pressure vector which is used to calculate $L_p(G)$. A paired image $X$ and real temperature field $Y$ are also used by $D$ to predict the label and the vapor pressure embedding which is used in calculating $L_p(D)$.

**Optical image to temperature mapping results**

The *G* and *D* architectures were adopted from previous works and modified for our application.[45, 47] Our input and output domains have different interpretations. The input was an array of pixel values showing the pixel intensities while the output was an array of temperatures depicting a physical parameter at every point. However, both follow a similar underlying structure. Therefore, it is reasonable to use architecture that helps with transferring the input structure to the output. Here, we used a U-Net[48] structure allowing for bypassing the information through skip connections, leveraging UNet inductive bias for domain translation. We employed least square GAN (LS-GAN)[49] loss to mitigate vanishing gradients. For the *D*, we used a convolutional PatchGAN which was introduced in previous work.[45] In a patchGAN, the discriminator will convolutionally move over $N \times N$ patches on the image to detect if the patch is from a real or fake output. We used a PatchGAN discriminator with a receptive field of $70 \times 70$ pixels. We modified the patchGAN by adding an additional head to output both class prediction (real or fake) and the pressure vector. The pressure embedding was concatenated with the input of *G*. Dropout with probability of 0.5 was used in the *G* network replacing noise *z* for stochasticity and to avoid overfitting.[45, 50]. More details on the networks and the training strategies are shown in the Methods section.

Training was done over a dataset of ~ 12000 paired grayscale images and temperature fields, including 87% hydrophobic samples and 13% superhydrophobic samples. The model was tested on 200 unseen data randomly chosen from different surfaces and vapor pressures. In all cases, the model was able to generate images with the correct structures with high temperature accuracy. Examples of generated temperature fields are shown in Fig. 7.

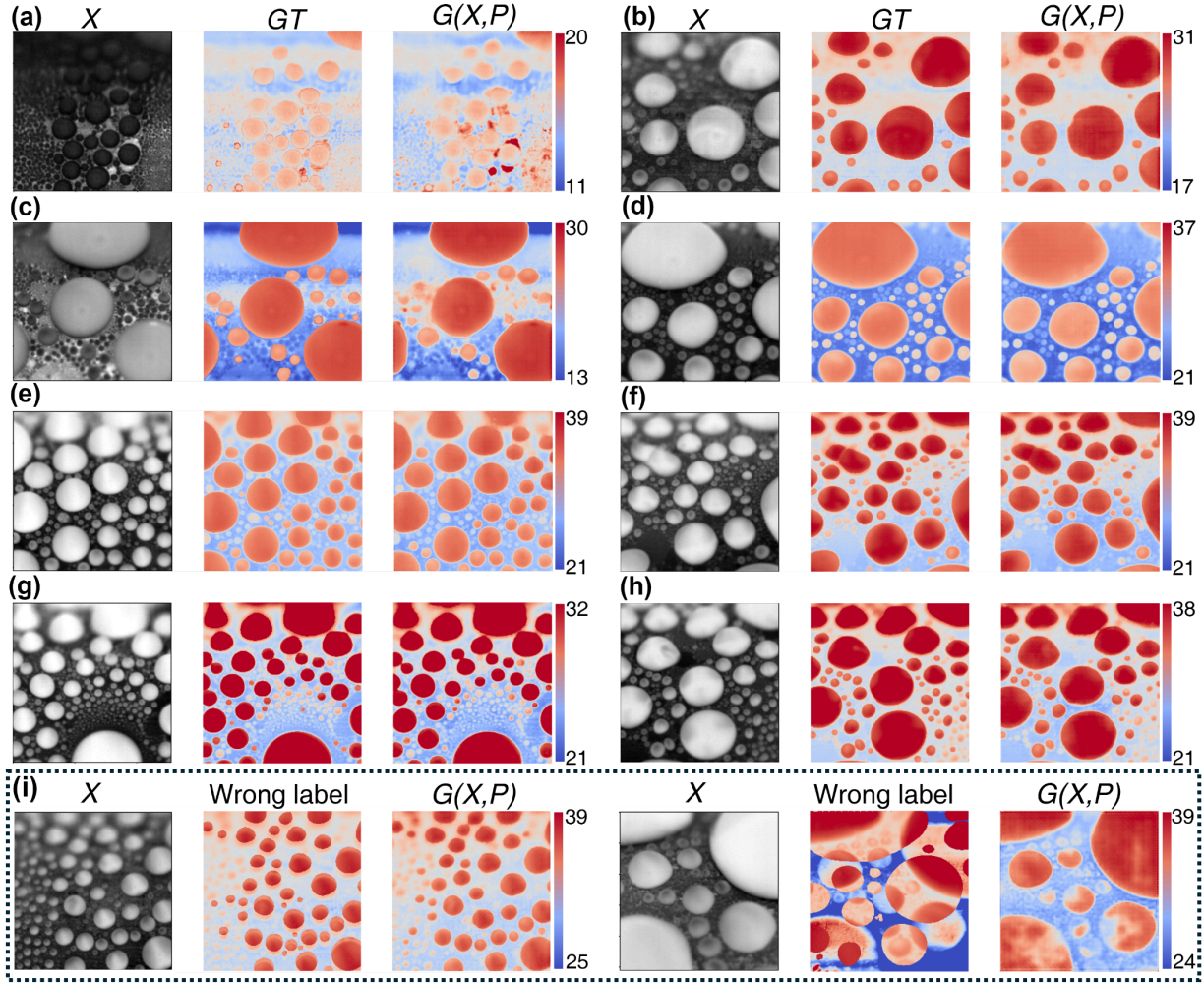

**Figure 7. Condensation optical image to temperature mapping**. Input data are randomly taken from the test dataset not seen by the model during the training or validation. (a) CuO with a Parylene C coating at $P$ = 2.6 kPa, (b) CuO with a Parylene C coating at $P$ = 4.5 kPa, (c) Cu with a PDMS coating at $P$ = 4.1 kPa, (d) CuO with a PDMS coating at $P$ = 5.4 kPa, (e) CuO with a Gentoo coating at $P$ = 6.4 kPa, (f) CuO with a PDMS coating at $P$ = 7.3 kPa, (g) CuO with a Gentoo coating at $P$ = 5.5 kPa, and (h) CuO with a PDMS coating at $P$ = 7.3 kPa. (i) Examples of images with wrong temperature labels in the training dataset. The left example in (i) shows an image with a slightly shifted label (minor deviation from correct distribution) while the right example in (i) shows an image with completely wrong labels (major deviation). Here, $X$ is in the gray scale image, $GT$ is the ground truth temperature map, and $G(X,P)$ is the model prediction. All temperatures are in $^0$C.

We evaluated the model performance using mean absolute error (MAE) between the predicted and real temperature fields. The MAE over 200 randomly chosen data from the training dataset, including both hydrophobic and superhydrophobic samples, was 0.57°C. The test dataset MAE was 1.3°C. Our analysis revealed that the generalization gap, where the model exhibits significantly higher errors on the test dataset compared to the training dataset, was primarily due to the superhydrophobic samples. When evaluated exclusively on the hydrophobic samples, the training and test errors dropped to 0.49°C and 0.57°C, respectively, demonstrating strong generalization. On the other hand, for superhydrophobic samples, the model appeared to overfit the training data rather than learning a robust mapping. We attributed this behavior to two reasons. First, the droplet detection model had lower accuracy for superhydrophobic samples due to the presence of smaller droplets departing the surface via coalescence-induced droplet jumping[36]. Second, the superhydrophobic samples constituted only 13% of the dataset which originated from a single condensation experiment. Consequently, the model learned the mapping based on majority hydrophobic samples, limiting its ability to generalize well to the underrepresented superhydrophobic samples. All subsequent evaluations are based on hydrophobic surfaces, unless otherwise stated.

We also evaluated the structural similarity between the real and generated samples using structural similarity index metric (SSIM), which is a common metric for evaluating synthetic image quality.[51] The model achieved an SSIM of 0.998 and 0.996 over training and test dataset, respectively, showing a strong resemblance between real and synthetic samples. While temperature accuracy holds greater importance over visual appearance, the model excelled at both. Interestingly, the model generated correct temperature maps even for the images with wrong labels in the training dataset (see Fig. 7i), suggesting that the model learned to infer droplet and substrate temperatures based on physical features such as droplet size, droplet morphology, location on the tube, and vapor pressure. This robustness indicated a strong learning, allowing the model to effectively disregard mislabeled data.

**Effect of vapor pressure conditioning and loss**

Conditioning of the GAN on the vapor pressure was essential to let the model learn different temperature distributions depending on the vapor pressure. The same droplet morphology and distribution may be observed on two different surfaces and at different vapor pressures. Without any vapor pressure information, there is no deterministic mapping solution for many of the samples, hindering the ability of the model to learn the correct mapping. To highlight the significance of vapor pressure conditioning, we trained the same GAN model on all the data at different vapor pressures without incorporating the vapor pressure conditioning step. In this case, the MAE over training and test datasets increased to 1.25°C and 1.5°C, showing a 155% and 163% increase, respectively. This result underscored the critical role of the conditioning step, ensuring the model effectively learns the relationship between the vapor pressure and temperature distribution. Additionally, incorporating the vapor pressure loss further enhances the model performance. When vapor pressure conditioning was applied without the vapor pressure loss term, the MAE increased to 0.62°C and 0.72°C for training and test dataset, respectively. This corresponds to approximately 26% prediction error increase for both train and test datasets. The benefit of introducing the pressure estimation loss is consistent with the domain classification loss proposed in multi-domain learning problem in a previous study.[46]. However, in our approach, we formulated this loss as a regression task instead of classification to comply with the continuous nature of the vapor pressure.

Finally, encoding the vapor pressure in a fixed embedding format resulted in better performance compared to the scalar format. When the vapor pressure information was given through a scalar value to the model, the MAE increased by 43% and 63% on the training and test datasets, respectively, demonstrating the significance of introducing the pressure information in an embedding format to the model. It should be noted that design and exhaustive optimization of the *G* and *D* networks was not conducted in this study. Our aim was to show the applicability of cGANs conditioned on a physical variable (vapor pressure) for temperature prediction during a phase change process, by combining adversarial loss, pointwise temperature loss, and pressure loss for improved accuracy and generalization. Future works can expand on our approach through additional network optimization and training strategies to achieve even lower prediction errors. A summary of all the quantitative results are shown in Table 1. Note that only the test dataset evaluations are shown in Table 1 (see Supplementary Table 4 for evaluation on training dataset). Visualization of the predictions by each model is shown in Fig. 8.

**Table 1. Quantitative results of image to temperature mapping.** The results are based on the model with no pressure conditioning (Model 1, $G(X)$), the model with pressure conditioning using a scalar $P_{vapor}$ and a pressure loss term (Model 2, $G(X, P_{scalar}) + P_{loss}$), the model with pressure conditioning using embedding of $P_{vapor}$ (Model 3, $G(X, P_{emb})$), and the model with pressure conditioning using embedding of $P_{vapor}$ and a pressure loss term (Model 4, $G(X, P_{emb}) + P_{loss}$). The results demonstrate the evaluation on a test dataset with both superhydrophobic (SHP) and hydrophobic (HP) samples and a test dataset with only HP samples.

|  | Model 1 | Model 2 | Model 3 | Model 4 |
|---|---|---|---|---|
| **Relative error, [%] (SHP + HP)** | 22.22 | 14.84 | 8.98 | **8.05** |
| **Relative error, [%] (HP)** | 10.15 | 7.23 | 4.98 | **4.26** |
| **MAE, [°K] (SHP + HP)** | 3.86 | 2.41 | 1.53 | **1.36** |
| **MAE, [°K] (HP)** | 1.45 | 0.93 | 0.72 | **0.57** |
| **SSIM (SHP + HP)** | 0.955 | 0.962 | 0.987 | **0.988** |
| **SSIM (HP)** | 0.986 | 0.984 | 0.995 | **0.996** |

Visualization of the predictions by each of the models is shown in Fig. 8. The model with vapor pressure embedding conditioning and an additional vapor pressure loss term (Model 4) performs the best. The significance of this model is more pronounced at lower vapor pressures where the other models over-estimate the temperatures in some regions.

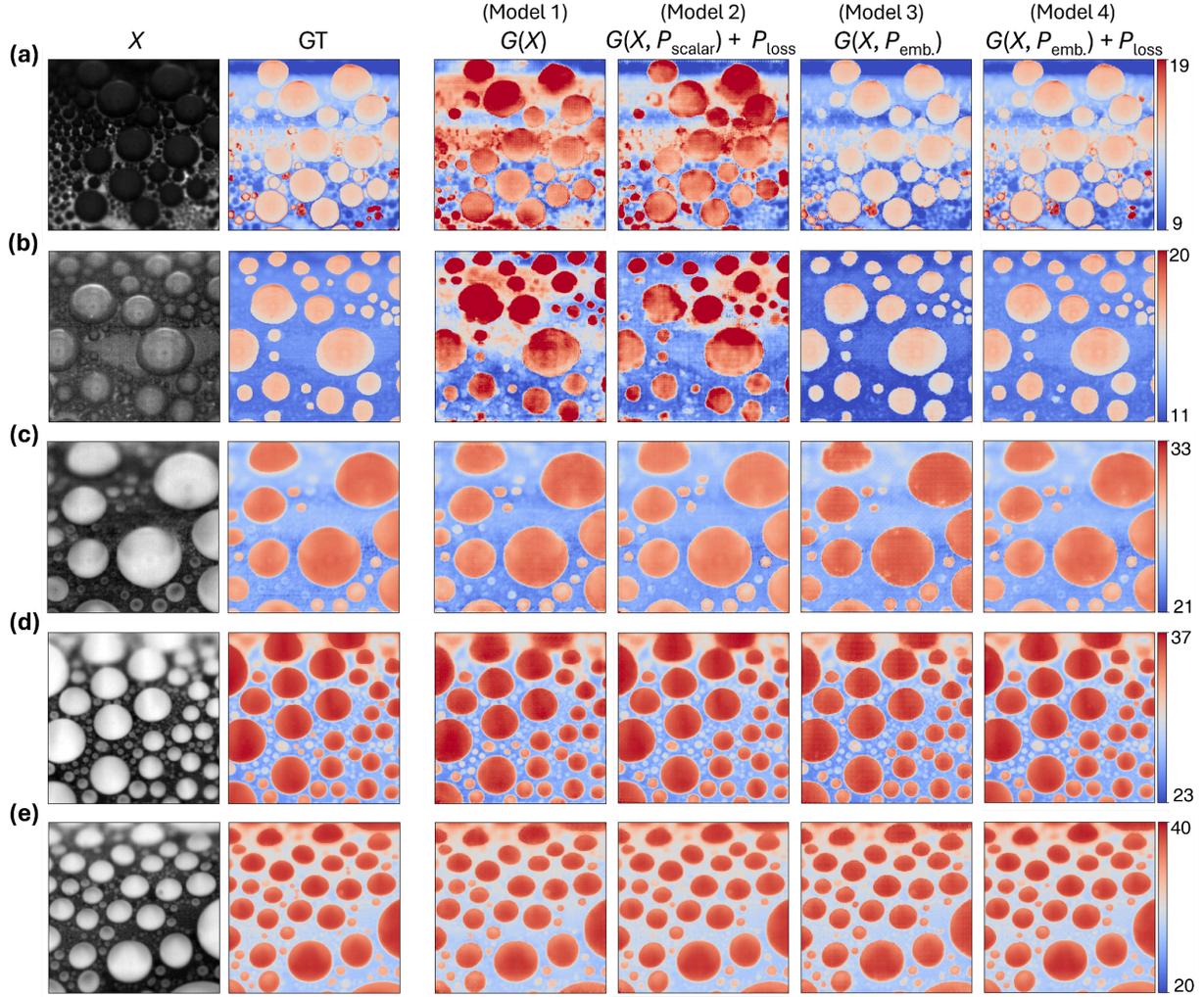

**Figure 8. Visualization of temperature predictions by each model (see Table 4).** (a) Cu with a PDMS coating at $P_{vapor}$ = 2.3 kPa, (b) CuO with a PDMS coating at $P_{vapor}$ = 2.3 kPa, (c) CuO with a Parylene C coating at $P_{vapor}$ = 4.5 kPa, (d) CuO with a Gentoo coating at $P_{vapor}$ = 6.4 kPa, (e) CuO with a PDMS coating at $P_{vapor}$ = 7.3 kPa. Here, $X$ and GT denote the input image and the ground truth temperature map, respectively. The models are without any pressure conditioning (Model 1, $G(X)$), with scalar pressure conditioning and a pressure loss (Model 2, $G(X, P_{scalar}) + P_{loss}$), with embedding pressure conditioning (Model 3, $G(X, P_{emb})$), and with embedding pressure conditioning and a pressure loss (Model 4, $G(X, P_{emb}) + P_{loss}$).

## Discussion

We developed a pixelwise calibration framework for IR thermography of pure steam condensation on surfaces with high spatial and temporal resolutions. This approach accounts for non-uniformities arising from two-phase interactions as well as inherent variations in the solid substrate. High-spatial-resolution measurements revealed the presence of an azimuthal temperature gradient on tubes during DWC. Statistical analysis of droplet dynamics further showed that larger droplets remain pinned in the upper region of the tube for extended periods. This behavior is primarily due to the smaller tangential gravitational forces acting on the droplets as well as the lower likelihood of sweeping interactions with other droplets in the upper region. These droplets remain at or near condensing vapor temperatures, resulting in the higher time-averaged temperature observed in this region. This finding was only made possible through the high-resolution IR thermography.

The local temperature patterns imply that the condensation heat transfer coefficient varies locally on the surface. The time-averaged accumulation of large droplets in the upper region resembles a local FWC mode, which is known for its poor heat transfer performance. Future studies can correlate these local temperature variations into local heat transfer coefficients, providing a new pathway for enhancing the design of condensation surfaces. The azimuthal temperature dependency reveals that the lower regions contribute more actively to the heat transfer. This enhanced contribution is due to the greater availability of nucleation sites, caused by faster droplet cycles in the lower regions compared to the upper ones. A droplet cycle is defined as a cycle including droplet nucleation, growth, and removal caused by either gravity force or sweeping effect of larger droplets. A droplet is removed from the surface when it grows large enough, known as critical size, that the gravitational force becomes larger than the droplet-surface adhesion force. Additionally, smaller droplets may be removed when they lie in the path of other larger droplets moving from top to bottom of the tube. In this process, larger droplets sweep away smaller droplets while moving downward (gravity direction) on the tube. However, the likelihood of smaller droplets removal increases toward the bottom of the tube due to the cumulative sweeping effect. These mechanisms collectively explain why the time-averaged droplet coverage and droplet size vary along the tube, leading to azimuthal temperature dependence.

While IR thermography combined with a rigorous calibration approach enables high-resolution temperature measurements during DWC, its complexity and cost have hindered widespread application in phase-change heat and mass transfer studies. This study develops an image-to-temperature mapping framework based on generative ML that marks a significant step toward eliminating the complex IR imaging and calibration bottlenecks for high-resolution temperature estimation on condensation surfaces. The proposed generative ML framework provides accurate, and high-resolution temperature prediction across various surfaces and vapor pressures. The $P_{vapor}$ information was incorporated into the model through a vapor pressure embedding at the input layer. Our results indicate that using an embedding format for this conditional parameter is more effective than simple scalar representation. Therefore, the developed GAN was conditioned on both the input image and the $P_{vapor}$ embedding during training and inference. To ensure the model predicts correct point-wise temperature values rather than just the overall distribution, as typically expected in GAN training, we included an $L_1$ term (Methods) in addition to the adversarial loss. Additionally, another loss term associated with the $P_{vapor}$ ($L_p$, Methods) penalizes both the discriminator and generator for incorrect vapor pressure prediction, thus inducing the importance of accurate vapor pressure estimation to the model. We showed that conditioning the model with the vapor pressure embedding is necessary for its generalization. Furthermore, incorporating the $L_p$ during the training also enhanced the predictions. When tested on hydrophobic surfaces, the model achieved relative errors of 2.9% and 4.2% on the training and test datasets, respectively, demonstrating high accuracy and strong generalizability. The errors increased to 5% and 8% when the superhydrophobic data were included in the evaluation. This performance drop is mainly due to the coalescence-driven jumping droplets, which remove smaller micro-scale droplets from the surface. These droplets are generally not well captured with the droplet segmentation model used in this study. Future works can address this challenge by introducing more superhydrophobic data and refining the droplet segmentation model, enabling the discovery of local temperature patterns and droplet statistics on superhydrophobic surfaces. While not explicitly tested on optical videos of condensation taken simultaneously with the IR imaging data, the learned mapping can be easily adopted to paired optical and temperature fields through transfer learning, significantly reducing costs and time by eliminating the need for expensive high-speed IR imaging systems and challenging calibration steps.

Finally, it is worth mentioning that our condensation IR thermography and calibration approach is also applicable to IR-transparent fluids. In this case, the droplet segmentation model and droplet temperature matrix are unnecessary, as the condensed liquid does not interfere with the IR signal emitted from the substrate. This enables direct temperature distribution measurement beneath the fluid layer during condensation, providing deeper insights into phase change dynamics and heat transfer. To take advantage of the developed image-to-temperature mapping model, transfer learning techniques can be applied to re-train the developed generative network with limited number of new data, enabling adaptation to new fluids.

**Methods**

**Condensation experimental facility**

The experimental facility comprises of a custom-designed stainless-steel pressure chamber with an internal diameter of 30.5 cm. This facility is vacuum compatible and can withstand internal pressure of up to 5.06 MPa. Tubes as long as 31 cm can be installed inside the chamber for condensation tests. The facility is equipped with several feedthroughs allowing for installation of thermocouples, resistance temperature detectors (RTDs), coolant lines, and pressure transducers. Two independent pressure transducers consisting of a MicroPirani and a Baratron 728A (both from MKS Instruments) were installed inside the chamber at two different locations to monitor the air and steam pressures within the chamber. Prior to each experiment, NCGs were removed from the chamber using a vacuum pump (Model Adixon Alcatel 2005). Condensation tests were started after the internal pressure of the chamber reaches below 5 Pa. A boiler filled with deionized (DI) water was heated using three tape heaters (Part #AWH-101-040DP, ETS Equipment) attached on the external surface of the boiler. The heating rate of these heaters was controlled by a variable voltage regulator (Model PM-1220B, ETS Equipment). The condenser surface was cooled down using a coolant flow loop connected to a large capacity chiller (Part #327005091602, System III TU7 Pump, Thermo Fisher Scientific). The coolant flow rate was measured using an electromagnetic flow meter (Part #FMG93, Omega). The coolant inlet and outlet temperatures were monitored using RTDs (Part #AT-PX1123Y-LR4S1T2T, ReoTemp). Except for the test section, all the other surfaces inside the chamber were thermally insulated to limit the condensation to the test section. Additionally, several tape heaters were wrapped around the chamber to heat the chamber walls prior to each experiment preventing condensation happening on the internal walls and viewports. Once the vacuum step was completed and the coolant inlet and outlet temperatures reached steady state conditions, the condensation test began by introducing the vapor into the chamber with the pressure set by adjusting the valve between the boiler and the chamber.

IR imaging of the condensation process was possible through a UV-grade sapphire glass viewport (Part #A0810-4-CF, MPF Products) installed on the side wall of the chamber (Fig. 1e). The UV-grade sapphire glass provides reasonably high transmission in the wavelength range of 0.17 – 5 µm, making it an appropriate option for IR imaging. The transmission of the sapphire glass is shown in Fig. 1d. A model X8503sc FLIR camera along with a 50 mm Macro MWIR FPO manual bayonet FLIR lens (Part #T300385) were used for imaging in the wavelength range of 3 – 5 μm.

**Droplet segmentation model**

The droplet segmentation model consisted of two parts: a droplet detection model based on Faster RCNN[52] to identify the approximate locations of the droplets on the surface using bounding boxes, and the Segment Anything Model (SAM) to specify the exact location of droplets. SAM, a vision foundational model developed by Meta, is designed to segment any object within an image directly during the inference step. However, we found out that SAM cannot directly segment the droplets on the surface unless the approximate location of each droplet is provided as the input prompts. To address this, we developed a droplet detection model based on Faster RCNN to detect the droplets approximate locations on the surface and use those locations as prompts to the SAM. The droplet detection model was trained on 400 manually labelled data and tested on another 70 data. Each image was cropped to a resolution of 520 × 640 pixels before train and test steps. The Faster RCNN was first initialized with pre-trained weights from ImageNet[53] dataset and re-trained with droplet images. Note that by integrating the Faster R-CNN predictions with SAM at inference stage, we avoided the need for training a full droplet segmentation model which saved a lot of computational cost for training and labor cost for training data preparation.

Evaluation of the droplet detection model is conducted by calculating the average precision (AP) and average recall (AR) of the detections (Supplementary Note 1). Once the approximate locations of the droplets were determined by the droplet detection model, the SAM was applied to accurately segment the droplets inside the detected regions. Notably, providing the approximate droplet locations was crucial for the performance of the SAM. Without these input prompts, droplet segmentation accuracy dropped significantly (< 0.1) and most of the droplets were not captured, highlighting the significance of the initial droplet detection stage

For training the droplet detection model, SGD optimizer with learning rate of 0.001 and momentum of 0.9 was used for optimization. The ResNet50 was used as the backbone of the droplet detection model. The droplet detection model achieved average precision and average recall of 0.95 when trained with batch size of 2 parallel on 2 GPUs for 150 epochs. These values were reduced to 0.61 and 0.5, when evaluated on test dataset (Supplementary Table 1 and 2).

**Generative machine learning model**

The original GAN training objective function is shown in Equation (1), where $G$ is the generator, $D$ is the discriminator, $x$ is the real data sampled from the distribution $p_{data}(x)$, z is the noise vector sampled from noise distribution $p_z(z)$. Here, $D(x)$ is the probability of having a real image expressed by the discriminator.

$$min_G \, max_D \, V(G, D) = E_{x \sim p_{data}(x)}[\log(D(x))] + E_{z \sim p_z(z)}[\log(1 - D(G(z)))] . \tag{1}$$

The loss function and objective function of our cGAN model can be expressed in the form of Equations (2-3), where $x$ and $y$ are the paired input data, and $p$ is the vapor pressure.

$$L_{cGAN}(G, D) = E_{x,y}[\log(D(x, y))] + E_{x,z}[\log(1 - D(x, G(x, z, p)))] , \tag{2}$$

$$G^* = min_G \, max_D \, L_{cGAN}(G, D) . \tag{3}$$

L1 loss ($L_{L1}$) was used in the training of the $G$ to penalize pointwise temperature deviations with the exact fields. $L_{L1}$ is shown in Equation (4).

$$L_{L1}(G) = E_{x,y,z}[\|G(x, z, p) - y\|_1] . \tag{4}$$

The vapor pressure loss functions for $D$ and $G$ are shown in Equations (5-6).

$$L_p(D) = E_{x,y}[\|D_p(x, y) - p\|_1] , \tag{5}$$

$$L_p(G) = E_{x,y,z}[\|D_p(x, G(x, z, p)) - p\|_1] . \tag{6}$$

The final loss function for the $G$ and $D$ can be expressed in the form of Equations (7-8), where $\lambda_{L1}$ and $\lambda_p$ are hyperparameters controlling the relative significance of pixel-level temperature prediction and vapor pressure estimation.

$$L_G = L_{cGAN}(G, D) + \lambda_{L1} L_{L1}(G) + \lambda_p L_p(G) \tag{7}$$

$$L_D = -L_{cGAN}(G, D) + \lambda_p L_p(D) \tag{8}$$

We observed more stable training with LS-GAN compared to vanilla GAN and Wasserstein GAN[54]. Therefore, we implemented a similar adversarial loss term as LS-GAN, which are expressed in Equations (9-10) for $G$ and $D$.

$$L_{cGAN}(G) = \tfrac{1}{2} E_{x,y,z}[(D(x, G(x, z, p)) - 1)^2] , \tag{9}$$

$$L_{cGAN}(D) = \frac{1}{2}E_{x,y}[(D(x, y) - 1)^2] + \frac{1}{2}E_{x,y,z}[(D(x, G(x, z, p)))^2], \quad (10)$$

**Network details and training strategies**

The *G* structure consists of eight downsampling steps in the encoder and eight upsampling steps in the decoder. Instance normalization, and leaky relu were used in *G* network. Instance normalization was used instead of batch normalization as it has been shown to be more useful for tasks like style transfer. In instance normalization, normalization is done across the spatial dimension of each individual sample.[55] During upsampling, modules of the leaky relu activation function, transpose convolution, and instance normalization were used. For the very last layer of decoder in U-Net, we used the softplus activation function. Softplus is a smooth approximation of the relu function and was used to constrain the output to positive values.[56] We explored using relu and tanh activation functions in the last layer and realized that zero or negative temperatures might be generated at several points of the output array which drastically impacts the accuracy. Using the softplus activation function introduced an inductive bias to enforce the model to create positive temperatures.

We used stochastic gradient descent (SGD) for training the *G* network and then the *D* network at every iteration. For the best model, $\lambda_{L1}$ and $\lambda_p$ were set to 150 and 0.1, respectively. The learning rate was initially set to 0.00075 and the model was trained for 40 epochs with a constant learning rate, and it was changed for the next 70 epochs using a cosine learning rate scheduler. All the trainings and experimentations were conducted on a HAL cluster at the national center for supercomputing applications (NCSA) at the University of Illinois at Urbana-Champaign.[57] Three GPUs of Nvidia Tesla V100 with 16 GB GPU memory and 5120 Cuda cores, and 32 CPU cores of IBM Power9 were used in parallel for training the model for 24 hours.

**Data, Materials, and Software Availability**

The codes, based on Pytorch implementation and parallel GPU computing, will be available on the first author Github repository: https://github.com/SiaK4/cGAN-image-to-temperature-mapping. A sample of the dataset will be included which can be used for testing the models. The full dataset used in this study will be available only upon reasonable request from the corresponding authors.

**Acknowledgments**

The authors gratefully acknowledge funding support from the Office of Naval Research (ONR) MURI under Grant No. N000142412575 with Dr. Mark Spector serving as the program officer. This work utilized computational resources at the National Center for Supercomputing Applications (NCSA) that are supported by the National Science Foundation's Major Research Instrumentation program, grant #1725729, as well as the University of Illinois at Urbana-Champaign. S.K. and N.M thank Dr. Kazi Fazle Rabbi and Dr. Jin Yao Ho for helping with the AM Al-alloy superhydrophobic sample preparation.

**Author contributions**

S.K. and N.M. conceived the idea for this research. S.K., C.W., and P.K. developed the infrared data collection method. S.K. developed the infrared thermography calibration framework. S.K. and P.K. carried out the condensation experiments. S.K. and T.S.T fabricated the samples. S.K. developed the image-to-temperature mapping models and the codes. S.K. and P.K. analyzed the infrared thermography data. S.K. and N.M. wrote the manuscript. All the authors have contributed to the paper and have given approval to the final version of the manuscript.

**Competing interests**

The authors have no competing interest to disclose.

# References

1. Dorsch, R.; Goodykoontz, J. *Local heat-transfer coefficients for condensation of steam in vertical downflow within a 5/8-inch-diameter tube*; 1966.
2. Kariya, K.; Sonoda, K.; Wakasugi, S.; Eshima, Y., Condensation and evaporation local heat transfer characteristics of the refrigerant mixture of R1123 and R32 inside a plate heat exchanger. **2018**.
3. Bergman, T. L.; Lavine, A. S.; Incropera, F. P.; DeWitt, D. P., *Introduction to heat transfer*. John Wiley & Sons: 2011.
4. Fu, W.; Chen, Y.; Inanlu, M. J.; Thukral, T. S.; Li, J.; Miljkovic, N., Enhanced pool boiling of refrigerants R-134a, R-1336mzz(Z) and R-1336mzz(E) on micro- and nanostructured tubes. *International Journal of Heat and Mass Transfer* **2024,** *220*, 124983.
5. Fazle Rabbi, K.; Khodakarami, S.; Ho, J. Y.; Hoque, M. J.; Miljkovic, N., Dynamic omniphobic surfaces enable the stable dropwise condensation of completely wetting refrigerants. *Nature Communications* **2025,** *16* (1), 1105.
6. Agonafer, D.; Spector, M. S.; Miljkovic, N., Materials and Interface Challenges in High-Vapor-Quality Two-Phase Flow Boiling Research. *IEEE Transactions on Components, Packaging and Manufacturing Technology* **2021,** *11* (10), 1583-1591.
7. Inanlu, M. J.; Ganesan, V.; Upot, N. V.; Wang, C.; Suo, Z.; Fazle Rabbi, K.; Kabirzadeh, P.; Bakhshi, A.; Fu, W.; Thukral, T. S.; Belosludtsev, V.; Li, J.; Miljkovic, N., Unveiling the fundamentals of flow boiling heat transfer enhancement on structured surfaces. *Science Advances* **2024,** *10* (45), eadp8632.
8. Köhler Mendizábal, J.; Singh, B. P.; Rabbi, K. F.; Upot, N. V.; Nawaz, K.; Jacobi, A.; Miljkovic, N., Enhanced internal condensation of R1233zd(E) on micro- and nanostructured copper and aluminum surfaces. *International Journal of Heat and Mass Transfer* **2023,** *207*, 124012.
9. Bucci, M.; Richenderfer, A.; Su, G.-Y.; McKrell, T.; Buongiorno, J., A mechanistic IR calibration technique for boiling heat transfer investigations. *International Journal of Multiphase Flow* **2016,** *83*, 115-127.
10. Gerardi, C.; Buongiorno, J.; Hu, L.-w.; McKrell, T., Infrared thermometry study of nanofluid pool boiling phenomena. *Nanoscale Research Letters* **2011,** *6* (1), 232.
11. Ganzevles, F. L. A.; van der Geld, C. W. M., Temperatures and the condensate heat resistance in dropwise condensation of multicomponent mixtures with inert gases. *International Journal of Heat and Mass Transfer* **2002,** *45* (15), 3233-3243.
12. Su, G. Y.; Wang, C.; Zhang, L.; Seong, J. H.; Kommajosyula, R.; Phillips, B.; Bucci, M., Investigation of flow boiling heat transfer and boiling crisis on a rough surface using infrared thermometry. *International Journal of Heat and Mass Transfer* **2020,** *160*, 120134.
13. Kim, S. H.; Lee, G. C.; Kang, J. Y.; Moriyama, K.; Park, H. S.; Kim, M. H., Heat flux partitioning analysis of pool boiling on micro structured surface using infrared visualization. *International Journal of Heat and Mass Transfer* **2016,** *102*, 756-765.
14. Bongarala, M.; Weibel, J. A.; Garimella, S. V., A method to partition boiling heat transfer mechanisms using synchronous through-substrate high-speed visual and infrared measurements. *International Journal of Heat and Mass Transfer* **2024,** *226*, 125516.
15. Wang, C.; Su, G.; Akinsulire, O.; Zhang, L.; Rahman, M. M.; Bucci, M., Investigation of critical heat flux enhancement on nanoengineered surfaces in pressurized subcooled flow boiling using infrared thermometry. *Heat Transfer Engineering* **2024,** *45* (4-5), 417-432.

16. Jung, J.; Kim, S. J.; Kim, J., Observations of the Critical Heat Flux Process During Pool Boiling of FC-72. *Journal of Heat Transfer* **2014,** *136* (4).
17. Mori, S.; Mt Aznam, S.; Yanagisawa, R.; Yokomatsu, F.; Okuyama, K., Measurement of a Heated Surface Temperature Using a High-Speed Infrared Camera During Critical Heat Flux Enhancement by a Honeycomb Porous Plate in a Saturated Pool Boiling of a Nanofluid. *Heat Transfer Engineering* **2020,** *41* (15-16), 1397-1413.
18. Bongarala, M.; Weibel, J. A.; Garimella, S. V., Boiling crisis is a consequence of thermal runaway in the substrate due to degradation of nucleate boiling heat transfer. *International Journal of Heat and Mass Transfer* **2024,** *233*, 125906.
19. Eimann, F.; Zheng, S.; Philipp, C.; Fieback, T.; Gross, U., Convective dropwise condensation out of humid air inside a horizontal channel – Experimental investigation of the condensate heat transfer resistance. *International Journal of Heat and Mass Transfer* **2018,** *127*, 448-464.
20. Ma, Z.; Ashraf, M. H.; Sun, M.; Zhang, G.; Wang, J., Investigation of condensate droplet movement in marangoni condensation of ethanol-water mixtures by infrared thermography. *Flow Measurement and Instrumentation* **2023,** *91*, 102373.
21. Kim, H.; Buongiorno, J., Detection of liquid–vapor–solid triple contact line in two-phase heat transfer phenomena using high-speed infrared thermometry. *International Journal of Multiphase Flow* **2011,** *37* (2), 166-172.
22. Guo, C.; Maynes, D.; Crockett, J.; Zhao, D., Heat transfer to bouncing droplets on superhydrophobic surfaces. *International Journal of Heat and Mass Transfer* **2019,** *137*, 857-867.
23. Kangude, P.; Srivastava, A., Understanding the growth mechanism of single vapor bubble on a hydrophobic surface: Experiments under nucleate pool boiling regime. *International Journal of Heat and Mass Transfer* **2020,** *154*, 119775.
24. Li, J.; Weisensee, P. B., Low Weber number droplet impact on heated hydrophobic surfaces. *Experimental Thermal and Fluid Science* **2022,** *130*, 110503.
25. Khodakarami, S.; Kabirzadeh, P.; Miljkovic, N., Self-supervised learning of shedding droplet dynamics during steam condensation. *APL Machine Learning* **2024,** *2* (2).
26. Suh, Y.; Chang, S.; Simadiris, P.; Inouye, T. B.; Hoque, M. J.; Khodakarami, S.; Kharangate, C.; Miljkovic, N.; Won, Y., VISION-iT: A Framework for Digitizing Bubbles and Droplets. *Energy and AI* **2024,** *15*, 100309.
27. Khodakarami, S.; Suh, Y.; Won, Y.; Miljkovic, N., Chapter Three - An intelligent strategy for phase change heat and mass transfer: Application of machine learning. In *Advances in Heat Transfer*, Abraham, J. P.; Gorman, J. M.; Minkowycz, W. J., Eds. Elsevier: 2023; Vol. 56, pp 113-168.
28. Goodfellow, I.; Pouget-Abadie, J.; Mirza, M.; Xu, B.; Warde-Farley, D.; Ozair, S.; Courville, A.; Bengio, Y., Generative adversarial nets. *Advances in neural information processing systems* **2014,** *27*.
29. Wang, J.; Zheng, G.; Yu, J.; Shao, J.; Zhou, Y., Sea Surface Temperature Prediction Method Based on Deep Generative Adversarial Network. *IEEE Journal of Selected Topics in Applied Earth Observations and Remote Sensing* **2024,** *17*, 14704-14711.
30. Jin, W.; Sadiqbatcha, S.; Zhang, J.; Tan, S. X.-D., Full-chip thermal map estimation for commercial multi-core CPUs with generative adversarial learning. In *Proceedings of the 39th International Conference on Computer-Aided Design*, Association for Computing Machinery: Virtual Event, USA, 2020; p Article 14.

31. Boroujeni, S. P. H.; Razi, A., IC-GAN: An Improved Conditional Generative Adversarial Network for RGB-to-IR image translation with applications to forest fire monitoring. *Expert Systems with Applications* **2024,** *238*, 121962.
32. Kang, M.; Phuong Nguyen, N.; Kwon, B., Deep learning model for rapid temperature map prediction in transient convection process using conditional generative adversarial networks. *Thermal Science and Engineering Progress* **2024,** *49*, 102477.
33. Mirza, M.; Osindero, S., Conditional generative adversarial nets. *arXiv preprint arXiv:1411.1784* **2014**.
34. Khodakarami, S.; Fazle Rabbi, K.; Suh, Y.; Won, Y.; Miljkovic, N., Machine learning enabled condensation heat transfer measurement. *International Journal of Heat and Mass Transfer* **2022,** *194*, 123016.
35. Chavan, S.; Cha, H.; Orejon, D.; Nawaz, K.; Singla, N.; Yeung, Y. F.; Park, D.; Kang, D. H.; Chang, Y.; Takata, Y.; Miljkovic, N., Heat Transfer through a Condensate Droplet on Hydrophobic and Nanostructured Superhydrophobic Surfaces. *Langmuir* **2016,** *32* (31), 7774-7787.
36. Ho, J. Y.; Rabbi, K. F.; Khodakarami, S.; Sett, S.; Wong, T. N.; Leong, K. C.; King, W. P.; Miljkovic, N., Ultrascalable Surface Structuring Strategy of Metal Additively Manufactured Materials for Enhanced Condensation. *Advanced Science* **2022,** *9* (24), 2104454.
37. Upot, N. V.; Fazle Rabbi, K.; Khodakarami, S.; Ho, J. Y.; Kohler Mendizabal, J.; Miljkovic, N., Advances in micro and nanoengineered surfaces for enhancing boiling and condensation heat transfer: a review. *Nanoscale Advances* **2023,** *5* (5), 1232-1270.
38. Rakrueangdet, K.; Nunak, N.; Suesut, T.; Sritham, E. In *Emissivity measurements of reflective materials using infrared thermography*, International MultiConference of Engineers and Computer Scientists (IMECS), 2016.
39. Zheng, S.-F.; Wu, Z.-Y.; Liu, G.-Q.; Yang, Y.-R.; Sundén, B.; Wang, X.-D., The condensation characteristics of individual droplets during dropwise condensation. *International Communications in Heat and Mass Transfer* **2022,** *131*, 105836.
40. Hu, X.; Yi, Q.; Kong, X.; Wang, J., A Review of Research on Dropwise Condensation Heat Transfer. *Applied Sciences* **2021,** *11* (4), 1553.
41. Upot, N. V.; Rabbi, K. F.; Khodakarami, S.; Ho, J. Y.; Mendizabal, J. K.; Miljkovic, N., Advances in micro and nanoengineered surfaces for enhancing boiling and condensation heat transfer: a review. *Nanoscale advances* **2023,** *5* (5), 1232-1270.
42. Li, L.; Khodakarami, S.; Yan, X.; Fazle Rabbi, K.; Gunay, A. A.; Stillwell, A.; Miljkovic, N., Enabling Renewable Energy Technologies in Harsh Climates with Ultra-Efficient Electro-Thermal Desnowing, Defrosting, and Deicing. *Advanced Functional Materials* **2022,** *32* (31), 2201521.
43. Ho, J. Y.; Fazle Rabbi, K.; Khodakarami, S.; Yan, X.; Li, L.; Wong, T. N.; Leong, K. C.; Miljkovic, N., Tunable and Robust Nanostructuring for Multifunctional Metal Additively Manufactured Interfaces. *Nano Letters* **2022,** *22* (7), 2650-2659.
44. Goodfellow, I.; Pouget-Abadie, J.; Mirza, M.; Xu, B.; Warde-Farley, D.; Ozair, S.; Courville, A.; Bengio, Y., Generative adversarial networks. *Communications of the ACM* **2020,** *63* (11), 139-144.
45. Isola, P.; Zhu, J.-Y.; Zhou, T.; Efros, A. A. In *Image-to-image translation with conditional adversarial networks*, Proceedings of the IEEE conference on computer vision and pattern recognition, 2017; pp 1125-1134.

46. Choi, Y.; Choi, M.; Kim, M.; Ha, J.-W.; Kim, S.; Choo, J. In *Stargan: Unified generative adversarial networks for multi-domain image-to-image translation*, Proceedings of the IEEE conference on computer vision and pattern recognition, 2018; pp 8789-8797.
47. Radford, A.; Metz, L.; Chintala, S., Unsupervised representation learning with deep convolutional generative adversarial networks. *arXiv preprint arXiv:1511.06434* **2015**.
48. Ronneberger, O.; Fischer, P.; Brox, T. In *U-net: Convolutional networks for biomedical image segmentation*, Medical image computing and computer-assisted intervention–MICCAI 2015: 18th international conference, Munich, Germany, October 5-9, 2015, proceedings, part III 18, Springer: 2015; pp 234-241.
49. Mao, X.; Li, Q.; Xie, H.; Lau, R. Y.; Wang, Z.; Paul Smolley, S. In *Least squares generative adversarial networks*, Proceedings of the IEEE international conference on computer vision, 2017; pp 2794-2802.
50. Mathieu, M.; Couprie, C.; LeCun, Y., Deep multi-scale video prediction beyond mean square error. *arXiv preprint arXiv:1511.05440* **2015**.
51. Guan, S.; Loew, M. In *Evaluation of Generative Adversarial Network Performance Based on Direct Analysis of Generated Images*, 2019 IEEE Applied Imagery Pattern Recognition Workshop (AIPR), 15-17 Oct. 2019; 2019; pp 1-5.
52. Girshick, R. In *Fast r-cnn*, Proceedings of the IEEE international conference on computer vision, 2015; pp 1440-1448.
53. Deng, J.; Dong, W.; Socher, R.; Li, L.-J.; Li, K.; Fei-Fei, L. In *Imagenet: A large-scale hierarchical image database*, 2009 IEEE conference on computer vision and pattern recognition, Ieee: 2009; pp 248-255.
54. Arjovsky, M.; Chintala, S.; Bottou, L. In *Wasserstein generative adversarial networks*, International conference on machine learning, PMLR: 2017; pp 214-223.
55. Ulyanov, D.; Vedaldi, A.; Lempitsky, V., Instance normalization: The missing ingredient for fast stylization. *arXiv preprint arXiv:1607.08022* **2016**.
56. Hao, Z.; Zhanlei, Y.; Wenju, L.; Jizhong, L.; Yanpeng, L. In *Improving deep neural networks using softplus units*, 2015 International Joint Conference on Neural Networks (IJCNN), 12-17 July 2015; 2015; pp 1-4.
57. Kindratenko, V.; Mu, D.; Zhan, Y.; Maloney, J.; Hashemi, S. H.; Rabe, B.; Xu, K.; Campbell, R.; Peng, J.; Gropp, W., HAL: Computer System for Scalable Deep Learning. In *Practice and Experience in Advanced Research Computing*, Association for Computing Machinery: Portland, OR, USA, 2020; pp 41–48.
58. Muley, S. V.; Ravindra, N. M., Emissivity of electronic materials, coatings, and structures. *Jom* **2014,** *66*, 616-636.
59. Thukral, T. S.; Rabbi, K. F.; Khodakarami, S.; Yang, W.; Sudarshan, A.; Pitschman, M. A.; Fourspring, P. M.; Miljkovic, N., Scalable and durable polydimethylsiloxane coating for anti-corrosion, anti-fouling, and condensation applications. *Cell Reports Physical Science* **2024,** *5* (10).
60. Khodakarami, S.; Zhao, H.; Rabbi, K. F.; Miljkovic, N., Scalable Corrosion-Resistant Coatings for Thermal Applications. *ACS Applied Materials & Interfaces* **2021,** *13* (3), 4519-4534.
61. Liu, K.; Vuckovac, M.; Latikka, M.; Huhtamäki, T.; Ras, R. H. A., Improving surface-wetting characterization. *Science* **2019,** *363* (6432), 1147-1148.

# Supplementary Information

**Supplementary Table 1**. Average precision (AP) of droplet detection model over training and test dataset.

|       | AP    | $AP_{small}$ | $AP_{medium}$ | $AP_{large}$ |
|-------|-------|--------------|---------------|--------------|
| Train | 0.948 | 0.931        | 0.990         | 0.999        |
| Test  | 0.611 | 0.455        | 0.728         | 0.902        |

**Supplementary Table 2**. Average recall (AR) of droplet detection model over training and test dataset.

|       | AR    | $AR_{small}$ | $AR_{medium}$ | $AR_{large}$ |
|-------|-------|--------------|---------------|--------------|
| Train | 0.954 | 0.938        | 0.992         | 0.999        |
| Test  | 0.651 | 0.512        | 0.765         | 0.915        |

**Supplementary Table 3.** Droplet advancing contact angles ($\theta_A$), receding contact angles ($\theta_R$), and contact angle hysteresis ($\Delta\theta = \theta_A - \theta_R$) on uniform 2 μm thick Parylene C, 1 μm thick PDMS, and 5 μm thick Gentoo coated surfaces, and the superhydrophobic (SHP) AM Al-alloy surface. Contact angle measurements were carried out using a microgoniometer (details in Section S4 of the Supporting Information). Five different measurements were taken on each surface, and the errors represent the standard deviation of the measurements.

| Coating        | $\theta_A$      | $\theta_R$      | $\Delta\theta$ |
|----------------|-----------------|-----------------|----------------|
| Parylene C     | 98.2 ± 1.3      | 73.3 ± 1.9      | 25 ± 2.3       |
| PDMS           | 111.2 ± 1.7     | 88.6 ± 5.5      | 22.6 ± 5.9     |
| Gentoo         | 104.6 ± 1.2     | 86.9 ± 4.6      | 17.1 ± 4.7     |
| SHP AM Al-alloy| 163.0 ± 3.0     | 161.3 ± 3.0     | 1.6 ± 4.2      |

**Supplementary Table 4. Evaluation of condensation image-to-temperature mapping models on training dataset.** The results are based on the model with no pressure conditioning (Model 1, $G(X)$), the model with pressure conditioning using scalar $P_{vapor}$ and a pressure loss term (Model 2, $G(X, P_{scalar}) + P_{loss}$), the model with pressure conditioning using embedding of $P_{vapor}$ (Model 3, $G(X, P_{emb})$), and the model with pressure conditioning using embedding of $P_{vapor}$ and a pressure loss term (Model 4, $G(X, P_{emb}) + P_{loss}$). The results demonstrate the evaluation on a dataset with both superhydrophobic (SHP) and hydrophobic (HP) samples and a dataset with only HP samples.

|                                | Model 1 | Model 2 | Model 3 | Model 4 |
|--------------------------------|---------|---------|---------|---------|
| Relative error, [%] (SHP + HP) | 7.37    | 5.25    | 4.98    | **4.08** |
| Relative error, [%] (HP)       | 7.09    | 4.84    | 4.27    | **2.89** |
| MAE, [°K] (SHP + HP)           | 1.22    | 0.77    | 0.71    | **0.40** |
| MAE, [°K] (HP) | 1.19 | 0.69 | 0.61 | **0.34** |
| SSIM (SHP + HP) | 0.997 | 0.990 | **0.997** | 0.995 |
| SSIM (HP) | **0.998** | 0.991 | 0.997 | 0.995 |

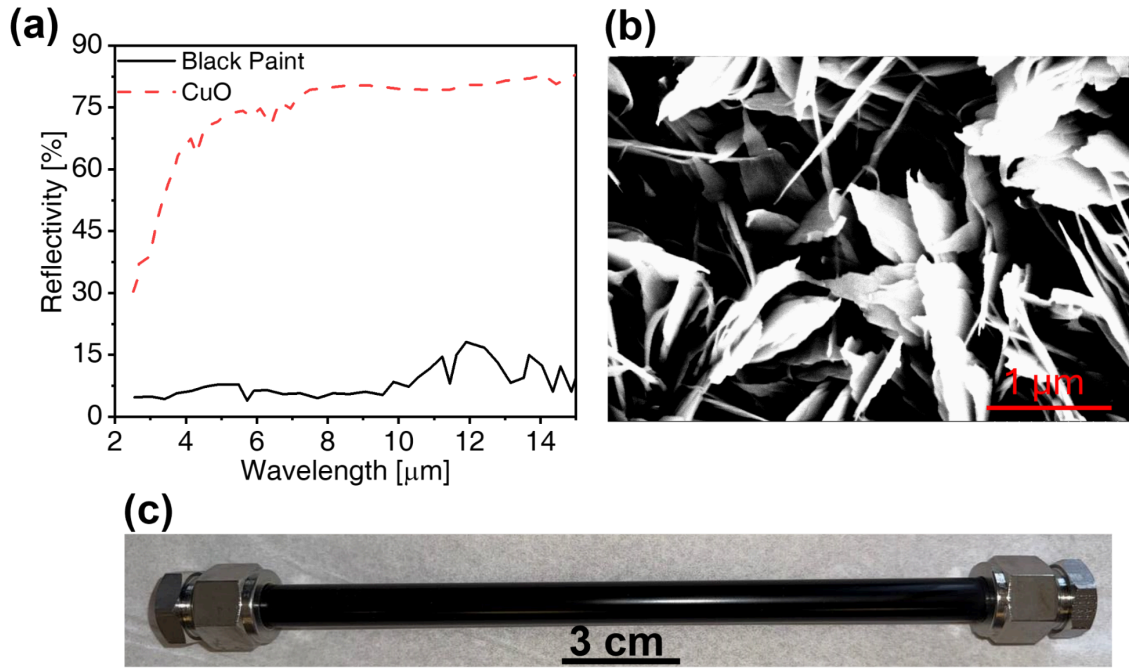

**Supplementary Figure 1. Details of CuO microstructures** (a) Measured reflectivity of the CuO compared with commercially available black paint (Rust-Oleum industrial paint). (b) Scanning electron microscopy (SEM) image of the CuO microstructures resulting in a visually black-like surface. (c) Optical image of a fabricated CuO coated tube sample.

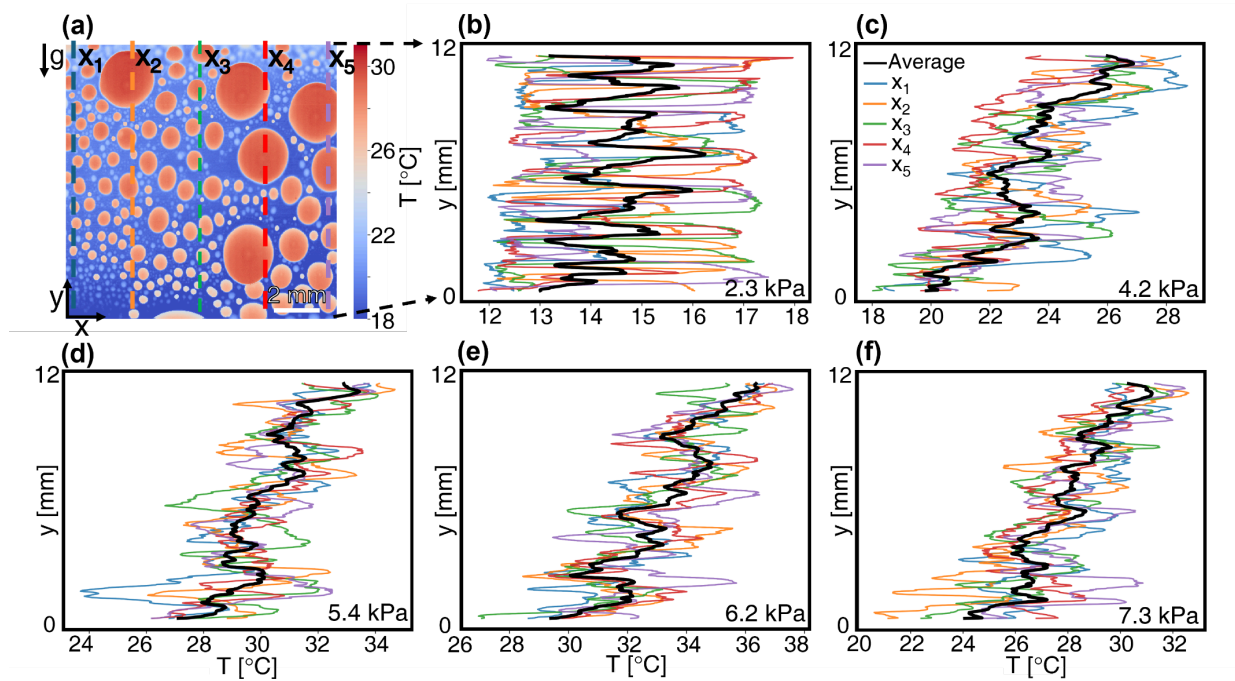

**Supplementary Figure 2. Time-averaged temperature variation in vertical direction (direction of gravity) during DWC on CuO with PDMS coating**. (a) Image showing the analyzed area with vertical lines ($x_1$ to $x_5$) used to visualize temperature variation. (b) Time-averaged vertical temperature variation at vapor pressure of 2.3 kPa. (c) Time-averaged vertical temperature variation at vapor pressure of 4.2 kPa. (d) Time-averaged vertical temperature variation at vapor pressure of 5.4 kPa. (e) Time-averaged vertical temperature variation at vapor pressure of 6.2 kPa. (f) Time-averaged vertical temperature variation at vapor pressure of 7.3 kPa. The black thicker line demonstrates average temperatures at 5 different lines of $x_1$ to $x_5$. Tube diameter is 12.7 mm (0.5"), however areas close to the bottom and top edges were out of focus and were removed for this analysis. The analyzed area has a surface area of 520 × 640 pixels or approximately 12 mm (height) × 14.7 mm (width). The legends in (c) are applicable to (a) and (d-f) figures as well.

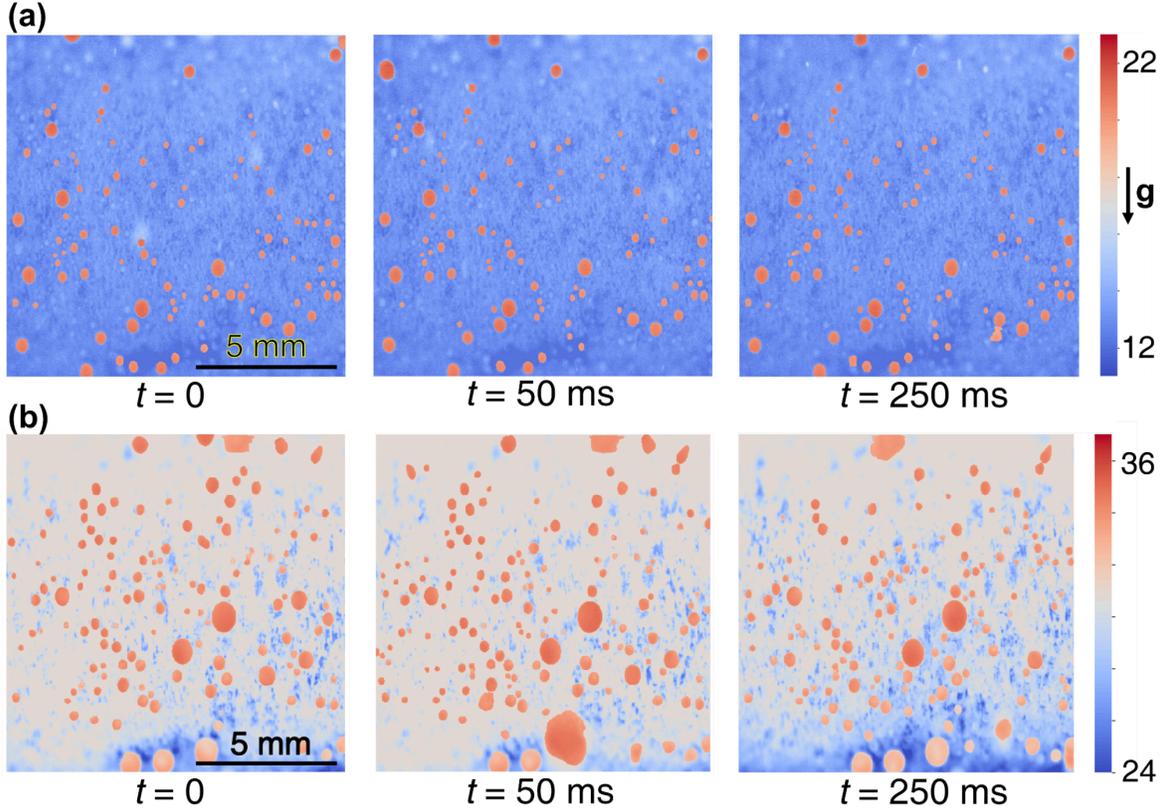

**Supplementary Figure 3. Transient temperature distribution of the superhydrophobic surface during condensation obtained through IR thermography**. Results showing condensation at vapor pressure of $P_{vapor}$ = 2.7 kPa, and (b) $P_{vapor}$ = 5.6 kPa. The superhydrophobic surface in both tests was SHP AM Al-alloy sample. Temperature units: $^0$C.

**Supplementary Notes**

**Supplementary Note 1:**

The total spectral radiation flux measured by the camera can be expressed by Equation (S1):

$$\phi_{\text{total}} = \int_{\lambda_1}^{\lambda_2} \tau_{\text{sapp},\lambda} \varepsilon_{\text{sur},\lambda} \phi_\lambda(T_{\text{sur}}) d\lambda + \int_{\lambda_1}^{\lambda_2} \varepsilon_{\text{c},\lambda} \tau_{\text{sapp},\lambda} (1 - \varepsilon_{\text{sur},\lambda}) \phi_\lambda(T_{\text{wall}}) d\lambda$$

$$+ \int_{\lambda_1}^{\lambda_2} \varepsilon_{\text{sapp},\lambda} \phi_\lambda(T_{\text{sapp}}) d\lambda + \int_{\lambda_1}^{\lambda_2} \varepsilon_{\text{bg},\lambda} \tau_{\text{sapp},\lambda}^2 (1 - \varepsilon_{\text{sur},\lambda}) \phi_\lambda(T_{\text{bg}}) d\lambda \quad \text{(S1)}$$

$$+ \int_{\lambda_1}^{\lambda_2} \varepsilon_{\text{bg},\lambda} \rho_{\text{sapp}} \phi_\lambda(T_{\text{bg}}) d\lambda + \int_{\lambda_1}^{\lambda_2} \varepsilon_{\text{air}} \phi_\lambda(T_{\text{air}}) d\lambda,$$

where $\tau_{sapp}$ is the sapphire glass transmissivity, $\varepsilon_{sur}$ is the test surface emissivity (the tube), $T_{sur}$ is the condenser surface temperature, $\varepsilon_c$ is the chamber wall emissivity, $T_{wall}$ is the chamber internal wall temperature, $\varepsilon_{sapp}$ is the sapphire glass emissivity, $T_{sapp}$ is the sapphire glass temperature, $\varepsilon_{bg}$ is the emissivity of background environment, $T_{bg}$ is the background environment temperature, $\rho_{sapp}$ is the sapphire glass reflectivity, $\varepsilon_{air}$ is the air emissivity, and $T_{air}$ is the ambient air temperature. Here, $\lambda_1$ and $\lambda_2$ are based on the IR camera wavelength working range. In this work, $\lambda_1 = 3$ μm and $\lambda_2 = 5$ μm. Furthermore, $\phi_\lambda(T)$ is the black body radiation based on Equation (S2), where $C_1$ and $C_2$ are constants, $C_1 = 3.742 \times 10^8$ (W·μm⁴)/m², and $C_2 = 1.439 \times 10^4$ μm·K.[3]

$$\phi_b(\lambda, T) = \frac{C_1}{\lambda^5 \left[\exp\exp\left(\frac{C_2}{\lambda T}\right) - 1\right]} . \tag{S2}$$

Radiation from ambient air is negligible. Similarly, the emissivity of sapphire is insignificant in the wavelength range smaller than 5 μm.[58] Therefore, the total radiation could be simplified into Equation (S3):

$$\phi_{total} \approx \int_{\lambda_1}^{\lambda_2} \tau_{sapp,\lambda} \varepsilon_{sur,\lambda} \phi_\lambda(T_{sur}) d\lambda + \int_{\lambda_1}^{\lambda_2} \varepsilon_{c,\lambda} \tau_{sapp,\lambda} (1 - \varepsilon_{sur,\lambda}) \phi_\lambda(T_{wall}) d\lambda$$

$$+ \int_{\lambda_1}^{\lambda_2} \varepsilon_{bg,\lambda} \tau_{sapp,\lambda}^2 (1 - \varepsilon_{sur,\lambda}) \phi_\lambda(T_{bg}) d\lambda$$

$$+ \int_{\lambda_1}^{\lambda_2} \varepsilon_{bg,\lambda} (1 - \tau_{sapp,\lambda}) \phi_\lambda(T_{bg}) d\lambda . \tag{S3}$$

**Supplementary Note 2: Precision, Recall, and Intersection over Union**

Precision is defined as the ratio of correctly predicted positive samples to the total number of predicted samples. It measures the accuracy of positive prediction. Recall, on the other hand, demonstrates the ability to detect all the positive cases and is measured as the ratio of predicted positive samples to the ratio of all the positive samples available in the data. In our case, high recall shows the ability to detect most of the droplets on the surface and high precision measures the accuracy of the detection. Precision and recall are defined by Equations (S4-S5), where TP is "True Positive", FP is "Falso Positive", and FN is "False Negative".

$$\text{Precision} = \frac{TP}{TP+FP} \tag{S4}$$

$$\text{Recall} = \frac{TP}{TP+FN} \tag{S5}$$

Intersection over union (IOU) was used as the metric to evaluate correct and incorrect predictions. In object detection, IoU measures the overlap between the predicted bounding box and the ground truth bounding box and is calculated as the ratio of overlap area between boxes to the union area of the two boxes. Ideally, the best model would achieve both high precision and recall. For each prediction, it is determined whether it is a true detection based on the IoU value. Generally, a lower IOU threshold increases the recall as it increases the likelihood of declaring a detection as positive. On the other hand, a higher IOU threshold results in higher precision. Mean average precision (mAP) is defined as the mean of average precision (AP) for different object classes. Since we only have one droplet class, mAP is equivalent to AP. Average recall (AR) is defined as the mean of recall values at different thresholds.

We evaluated the droplet segmentation accuracies depending on the droplet size by categorizing the droplets into small, medium, and large droplets. Small droplets were defined as droplets with area coverage smaller than $32 \times 32$ pixels, medium droplets were those with a coverage area in range of [$32 \times 32$, $96 \times 96$] pixels, and large droplets had area coverages more than $96 \times 96$ pixels.

The droplet detection model achieved AP of 0.95 and 0.61, and AR of 0.95 and 0.65 over training and test datasets, respectively. The detection accuracies dropped for smaller droplets compared to larger droplets (Supplementary Table 1 and 2).

**Supplementary Note 3: Surface Fabrication**

PDMS coating solution was prepared by Two-part Sylgard™ 184 Silicone Elastomer kit (Dow Corporation). The kit consists of a base (ethylbenzene, CAS#: 100-41-4) and a curing agent (methylvinylcyclosiloxane, CAS#: 2554-06-5). To obtain a 1 micrometer thick coating, 9.1 grams of the elastomer base and 130 grams n-hexane (Sigma Aldrich, CAS#: 110-54-3) in a glass container by stirring for 5 minutes, followed by addition of 0.91 grams of the curing agent, and further stirring for 5 minutes. Then, the solution was allowed to stand for 5 minutes to get rid of air bubbles created due to stirring.[59]

The Gentoo™ coating kit includes two components (parts A and B). The coating was prepared according to the manufacturer application guide. Briefly, the two coating components (Part A and B) were mixed (equal weights) and continuously stirred for 60 minutes using a magnetic stir bar to allow complete hydrolysis. The cleaned tubes were then dipped into the coating solution for 30 seconds and withdrawn at a constant speed of ~5 mm/second. The tubes were left to dry in air for 10 minutes, followed by curing at 100 °C for PDMS and 90 °C for Gentoo™, respectively, for 60 minutes in a closed atmospheric pressure furnace (SVAC2, Shel Lab). The curing times were chosen based on the manufacturer's application guides.

Parylene C was deposited through a chemical vapor deposition (CVD) process using a Specialty Coating Systems Labcoater 2 deposition system, resulting in a conformal hydrophobic coating.[60] Solid Parylene-C dimer particles (Chlorinated poly(para-xylylene)) were placed in the deposition system and sublimated at 150°C in vacuum conditions. The dimeric Parylene gas was cleaved to its monomeric form at 690°C, which was then condensed and polymerized as a conformal coating on the tube. The thickness of the coating can be controlled by varying the amount of dimer and the deposition time. In this study, the thickness was set to 2 μm to ensure coating uniformity while not degrading the thermal performance caused by the large thermal resistance of the Parylene layer.

The CuO microstructure was created by immersing the tubes in an alkaline solution composed of sodium chlorite, sodium hydroxide, and sodium phosphate at 95°C for 10 minutes. The Cu tubes were cleaned by a 2 M hydrochloric acid for 30 seconds, rinsed with DI water, and dried with a dry nitrogen gas stream prior to the immersion in the alkaline solution. Figure C1 demonstrates the scanning electron microscopy (SEM) images of the CuO surface and the measured reflectivity.

The AM Al-alloy tube was fabricated based on the recipe in previous work.[36] AlSi10Mg alloy was selected as the base materials and the AM samples were fabricated using the SLM280HL AM facility. To generate 3D structures that enhance the surface superhydrophobicity, the AM sample was further etched with a 2.0 M hydrochloric acid (HCl). The etching step will preferentially remove the Al-rich phase but the Si-rich cell walls, generating a 3D cellular structures. Finally, the etched AM sample was boehmitized to form a second-tier nanostructures on the surface. The boehmitization was carried out by immersing the sample in a boiling DI water for 30 minutes. A very thin layer (~ 3 nm) of heptadecafluoro-1,1,2,2-tetrahydrodecyl trimethoxysilane (HTMS) was deposited on the AM Al-alloy to render the surface superhydrophobic. HTMS is a self-assembled monolayer (SAM) which was deposited through atmospheric pressure CVD at 100°C for elongated time (~ 24 hours). To achieve HTMS deposition, 5% v/v HTMS was mixed with toluene and placed in a clean glass container. The mixture was then placed in a larger container along with the AM Al-alloy samples. The larger container was then sealed and placed in an atmospheric pressure oven (Thermo Scientific BF51732C-1).

All the coatings were applied on tubes with outer diameter of $d_o$ = 12.7 mm (0.5"), inner diameter of $d_i$ = 11 mm (0.436"), and length of 25.4 cm (10").

**Supplementary Note 4: Contact Angle Measurement**

Contact angle measurements were carried out using microgoniometry (MCA 3, Kyowa Interface Science). The apparent advancing contact angle ($\theta_a$), apparent receding contact angle ($\theta_r$) and contact angle hysteresis ($\Delta\theta = \theta_A - \theta_B$) were measured on four different spots for each sample and to gain understanding of measurement uncertainty. Note, measurement uncertainty may be larger than the reported standard deviation of the multiple measurements.[61] All contact angle data were analyzed using the image processing software (FAMAS, Interface Measurement and Analysis System) with the circle fitting method.